\PassOptionsToPackage{usenames, dvipsnames}{color}
\documentclass[10pt,journal,compsoc]{IEEEtran}
\ifCLASSOPTIONcompsoc
  \usepackage[nocompress]{cite}
\else
  \usepackage{cite}
\fi

\ifCLASSINFOpdf
\else
\fi

\usepackage{amsmath,amsthm,amssymb,amsfonts,bm}
\newtheorem{theorem}{Theorem}

\newtheorem{lemma}{Lemma}
\newtheorem{problem}{Problem}
\newtheorem{proposition}{Proposition}
\newtheorem{game}{Sub-Game}
\newtheorem{question}{Question}

\usepackage{stfloats}
\usepackage{cases}
\newtheorem{defn}{Definition}
\usepackage{algorithmic}
\usepackage{graphicx}
\usepackage{epstopdf}
\usepackage{textcomp}
\usepackage{xcolor}
\usepackage{url}
\usepackage{enumitem}
\usepackage{url}
\usepackage{bbm}
\usepackage{ragged2e}
\setlist[itemize]{leftmargin=*}
\setlist[enumerate]{leftmargin=*}
\usepackage{multirow}
\newcommand{\tabincell}[2]{\begin{tabular}{@{}#1@{}}#2\end{tabular}} 
\usepackage{diagbox}
\usepackage{subfigure}
\usepackage[linesnumbered, ruled]{algorithm2e}
\usepackage{makecell}
\usepackage{textcomp,booktabs}

\usepackage[usenames,dvipsnames]{color}
\usepackage{colortbl}
\definecolor{mygray}{gray}{.9}
\definecolor{mypink}{rgb}{.99,.91,.95}
\definecolor{mycyan}{cmyk}{.3,0,0,0}
\definecolor{myblue}{rgb}{.8,0.98,.9}
\begin{document}
	
	\title{Incentivized Federated Learning and Unlearning}
	\author{Ningning Ding, Zhenyu Sun, Ermin Wei, and Randall Berry,~\IEEEmembership{Fellow,~IEEE}
			
		\IEEEcompsocitemizethanks{\IEEEcompsocthanksitem Ningning Ding, Zhenyu Sun, and Randall Berry are with the Department of Electrical and Computer Engineering, Northwestern University, Evanston, IL 60208, USA (email: ningning.ding@northwestern.edu). 
			\IEEEcompsocthanksitem Ermin Wei is with Electrical and Computer Engineering Department and Industrial Engineering and Management Sciences Department, Northwestern University, Evanston, IL 60208, USA. 
		} 
	}

\IEEEtitleabstractindextext{%
\begin{abstract}
\justifying	To protect users'  \emph{right to be forgotten}  in federated learning, federated unlearning aims at eliminating the impact of leaving users'  data  on the global learned  model. 
The current research in federated unlearning mainly concentrated on developing effective and efficient unlearning techniques. However, the issue of incentivizing valuable users to remain engaged and preventing their data from being unlearned is still under-explored, yet important to the unlearned model performance. 
This paper focuses on the incentive issue and develops an incentive mechanism for federated learning and unlearning. We first characterize the leaving users' impact on the global model accuracy and the required communication rounds for unlearning.
Building on these results, we propose a four-stage game to capture the interaction and information updates during the learning and unlearning process. A key contribution is to summarize users’ multi-dimensional private information into one-dimensional metrics to guide the incentive design.  
We further investigate whether allowing federated
unlearning is beneficial to the server and users, compared to a scenario without unlearning. Interestingly,
users usually have a larger total payoff in the scenario with higher costs, due to the server’s excess incentives
under information asymmetry. 
The numerical results demonstrate the necessity of unlearning incentives for retaining valuable leaving users,  and also show that our proposed mechanisms decrease the server's cost by up to 53.91\% compared to state-of-the-art benchmarks.
\end{abstract}

}

\maketitle

\IEEEdisplaynontitleabstractindextext

\IEEEpeerreviewmaketitle

\IEEEraisesectionheading{\section{Introduction}\label{sec:introduction}} 
\subsection{Background and Motivations}
\IEEEPARstart{F}{ederated} learning is a promising distributed machine learning paradigm, in which multiple users collaborate to train a shared model under the coordination of a central server \cite{mcmahan2016communication}. This approach allows users to keep their local data on their own devices and only share the intermediate model parameters, which helps protect their raw data. However, despite these measures, it may not provide sufficient privacy guarantees \cite{truex2019hybrid,mothukuri2021survey}.

For privacy reasons, one desirable property of a federated learning platform is the users' ``right to be forgotten'' (RTBF), which has been explicitly stated in the European Union General Data Protection Regulation (GDPR) \cite{voigt2017eu} and the California Consumer Privacy Act (CCPA) \cite{harding2019understanding}. 
That is, a user has the right to request deletion of his private data and its impact on the trained model, if he no longer desires to participate in the platform. 
Users may seek to leave a platform for a variety of reasons.  For example, they may feel that the benefits from the platform are not sufficient to compensate for their potential privacy leakage from participation. Furthermore, until they participate in the platform, they may not have full knowledge of these benefits and costs due to incomplete information about other
users’ data. For instance, users'   privacy costs in federated learning depend on how unique their data is \cite{jiang2019improving}, which they can infer from their training loss after training \cite{gao2022verifi}.


To remove data from a trained federated learning model, the concept of \emph{federated unlearning}  has recently been proposed \cite{liu2020learn}. In this concept, after some users request to revoke their data, staying users will perform additional training or calculations to eliminate the impact of leaving users' data and obtain an unlearned model.  A simple yet costly approach is to retrain the model from scratch with the requested data being removed from the training dataset \cite{bourtoule2021machine}. To be more efficient and effective, existing literature (e.g., \cite{liu2021federaser,gao2022verifi,wu2022federated}) focused on alternative federated unlearning methods that obtain a model similar (in some distance metrics) to a retrained model with lower computational costs. 
However, these studies usually assumed that users are willing to participate in federated learning and unlearning. This assumption may not be realistic without proper incentives since users incur various costs during the training process (e.g., time, energy, and privacy costs). Our goal in this paper is to develop incentive mechanisms to help retain valuable leaving users and create a sustainable learning platform for both the users and the server. 

To design the incentive mechanism for federated learning and unlearning, there are several challenges to tackle.  
First, different leaving users will lead to different unlearned model performances and unlearning costs, the relationship among which is still an open problem yet essential for designing incentives. 
Second,  it is difficult for the server to design incentives for a large number of heterogeneous users, when users have multi-dimensional private information (e.g., training costs and privacy costs) and unknown information (e.g., users' training losses before federated learning).  
Third,   unlearning incentives for retaining valuable leaving users require careful design. High incentives may encourage strategic users to intentionally request revocation to obtain retention rewards, while low incentives may fail to retain valuable users. It is also crucial for the server to distinguish between high-quality leaving users (e.g., with rare and valuable data) and low-quality ones (e.g., with erroneous data),  both of which can lead to high training losses. 
Fourth,  both learning and unlearning incentives affect the server's and users' payoffs but are determined in different stages - before or after federated learning. Meanwhile, there are different information asymmetry levels in each stage, as the federated learning process can reveal some information such as users' training losses and contributions.   Thus, the mutual influence of the incentives and dynamic information asymmetry further complicate the incentive mechanism design.  

The above discussion motivates us to answer the following interesting question: 
\begin{question}
	Considering leaving users' impact, what is the server's optimal incentive mechanism for federated learning and unlearning,  when heterogeneous users have strategic data revocation decisions and multi-dimensional private and unknown information?
\end{question}

Furthermore, although federated unlearning is important   for protecting users' right to be forgotten and data privacy,   no work has studied whether allowing  federated unlearning is economically beneficial to the server or users by comparing the following two scenarios:
\begin{itemize}
	\item \emph{Unlearning-Allowed Scenario.} The federated learning server allows users to revoke data and will perform federated unlearning;
	
	\item \emph{Unlearning-Forbidden Scenario.} The federated learning server does not allow users to revoke data after they decide to participate in the federated training.	
\end{itemize}
Different unlearning scenarios will lead to different optimal incentive mechanisms, as well as the server's and users' payoffs. When unlearning is optional, studying the superiority of each scenario will facilitate the server's and users' selection. The performance comparison also provides insights into the policy design of a market regulator. 
This motivates the second key question of this paper:
\begin{question}
	Compared with the unlearning-forbidden scenario, is the federated unlearning-allowed scenario more beneficial to the server and users in terms of their payoffs?
\end{question}

\subsection{Contributions}
We summarize our key contributions below.
\begin{itemize}
	\item \emph{Incentive mechanism design for federated learning and unlearning.}  
	We propose a four-stage Stackelberg game to analyze the optimal incentives of the server and the optimal strategies of users within this game. 
	To the best of our knowledge, this is the first analytical study of incentive mechanisms for federated learning and unlearning. 
	\item \emph{Theoretical characterization of global model accuracy and unlearning communication rounds.} 
	We theoretically derive bounds on the global model optimality gap given non-IID data for   federated learning algorithms (Scaffold \cite{karimireddy2020scaffold} and FedAvg \cite{mcmahan2016communication}) and the number of global communication rounds required for a federated unlearning method. 
	\item \emph{Optimal incentives and revocation decisions under multi-dimensional incomplete information.} 
	Due to the complex interaction, users' multi-dimensional private information, and dynamically updated knowledge, the server's optimization problem in Stage I of the four-stage game is highly complex. 
	We	summarize users' multi-dimensional heterogeneity into several one-dimensional metrics and develop an efficient algorithm with linear complexity,  to handle the exponentially large number of possible cases involved in optimal mechanism design. 
	We also identify and analyze a supermodular game among the users to obtain their optimal data revocation decisions. 
		\item \emph{Comparison of unlearning-allowed and unlearning-forbidden scenarios.} 	We show that (i) when users' unlearning costs in the unlearning-allowed scenario are large, the server needs to compensate them with large incentives and thus prefers the unlearning-forbidden scenario. Surprisingly, users prefer the unlearning-allowed scenario where they have large costs, due to the excess rewards they obtain under information asymmetry. 		(ii) When users' perceived privacy costs in the unlearning-forbidden scenario are large, the server prefers the unlearning-allowed scenario while users prefer the unlearning-forbidden scenario for similar reasons as in (i).
	\item \emph{Insights and Performance Evaluation.} 
	We show that high costs and training losses motivate users to leave, while the server will retain the leaving users who make significant contributions to model accuracy but not necessarily low training losses, as small losses of retained users will reduce privacy costs yet increase unlearning costs. 
	We numerically show that compared with state-of-the-art benchmarks, our proposed incentive mechanism decreases
	the server's cost by up to 53.91\%. Moreover, the results demonstrate that it is beneficial for the server to retain valuable leaving users and jointly optimize the federated learning and unlearning incentive mechanisms.
\end{itemize}

\subsection{Related Work}
The concept of \emph{machine unlearning}, which refers to the process of removing the impact of a data sample from a trained model, was first introduced by Cao et al. in 2015 \cite{cao2015towards}. 
Most related literature was about centralized machine unlearning (e.g., \cite{bourtoule2021machine,ginart2019making}), in which the unlearned model (not retrained from scratch) was trained on summarized (e.g., aggregates of summations) or partitioned subsets rather than individual training samples. As a result,  the model only needed to be updated on the subset(s) of data that are associated with the requested samples.

Centralized unlearning methods are not suited to federated learning, due to (i) lack of direct data access, (ii) the fact that the global
model is updated based on the aggregated rather than the raw gradients, and (iii) the possibility that different users may have similar training samples \cite{gao2022verifi}. This motivated the emergence of \emph{federated unlearning}, which focuses on deleting the impact of revoked data in federated learning.

Only a few studies proposed federated unlearning mechanisms using methods such as gradient subtraction (e.g., \cite{liu2021federaser,liu2020learn}),   gradient scaling (e.g., \cite{gao2022verifi}), or knowledge distillation  (e.g., \cite{wu2022federated}).  Albeit with good numerical performance, there is no theoretical guarantee of these proposed federated unlearning methods. To fill this gap, we propose theoretical bounds on the model optimality gap and communication rounds for one approach to federated unlearning in this paper.

Furthermore, there is a wide spectrum of literature on incentive mechanisms for various systems, including crowdsensing (e.g., \cite{xie2014modeling}), wireless networks (e.g., \cite{zhao2018dynamic}),   data trading (e.g., \cite{wang2016value}), and energy sharing (e.g., \cite{wang2019incentive}). 
Some important work studied incentive mechanism design for federated learning to discourage valuable clients from leaving (e.g., \cite{zhan2020learning,ding2020optimal,zhang2021faithful,wang2022socially}). However, very few of them considered users' multi-dimensional private information (e.g., \cite{ding2020optimal}), and none of them incorporated the unique aspects of federated unlearning (e.g., unlearning costs) or the  dynamics of users’
payoffs (e.g., pre-/post-training and before/after some users leave). This paper is the first to focus on incentive mechanism design for both federated learning and unlearning. 

The rest of the paper is organized as follows. In Section \ref{chara}, we characterize the models of federated learning and unlearning. The system model is described in Section \ref{system}. We give the optimal incentive mechanisms in the unlearning-allowed and unlearning-forbidden scenarios in Sections  \ref{allowed}  and \ref{forbidden}, respectively. We provide simulation results in Section \ref{simulation} and conclude in Section \ref{conclusion}.

\section{Characterization of Federated Learning and Unlearning Models}
\label{chara}
Before modeling the game-theoretic interaction between the server and the users in the next section,   we first discuss federated learning and unlearning models in this section as a preliminary. Specifically, we specify the learning and unlearning objectives in Sections \ref{learningo} and \ref{unlearningo}, respectively. Then, we derive bounds on global model accuracy and federated unlearning time in Section \ref{bounds}. 
\subsection{Federated Learning Objective}
\label{learningo}
Consider an example of data $(x_a,y_a)$, where $x_a$ is the input (e.g., an image) and $y_a$ is the label (e.g., the object in the image). The objective of learning is to find the proper model parameter $w$ that can predict the label $y_a$ based on the input $x_a$. Let us denote the prediction value as $\tilde{y}(x_a;w)$. The gap between the prediction  $\tilde{y}(x_a;w)$ and the ground truth label $y_a$ is characterized by the prediction loss function $f_a(w)$. 
If user $i$ selects a set  of local data with data size $d_i$ to train the model,  the loss function of user $i \in \mathcal{I}$  is the average prediction loss on all his training data: 
\begin{equation}
F_{i}(w)=\frac{1}{d_{i}} \sum_{a=1}^{d_i} f_{a}(w).
\end{equation}
The purpose of federated learning is to compute the model parameter $w$ by using all users' local data. The optimal model parameter $w^*$  minimizes the global loss function, which is an average of all users' loss functions \cite{karimireddy2020scaffold,pathak2020fedsplit}:\footnote{This model treats each user equally. Some papers (e.g., \cite{mcmahan2016communication}) adopted another objective, a weighted sum of all users' losses, where the weights (i.e. $d_i/\sum_{i=1}^Id_i$) reflect the differences in data size. The two objectives are equivalent when users' data sizes are the same. Our results can be easily extended to the weighted case.}
\begin{equation}
\label{weight}
w^*=\arg\min _{w}  F(w)\triangleq\arg\min _{w}\frac{1}{I}\sum_{i\in \mathcal{I}} F_{i}(w).
\end{equation}

%


\subsection{Federated Unlearning Objective}
\label{unlearningo}
A federated learning process maps users' data into a model space, while a federated unlearning process maps a  learned model,  users' data set, and the data set that is required to be forgotten into an unlearned model space. The goal of federated unlearning is to make the unlearned model have the same distribution as the retrained model (i.e., retrained from scratch using the remaining data).\footnote{The distribution is due to the randomness in the training process (e.g. randomly sampled data and random ordering of batches).}


A natural method for federated unlearning  is to let the remaining users (excluding leaving users) continue training  from the learned model $w^*$, until it converges to  a new optimal model parameter $\tilde{w}^*$ that minimizes the global loss function of remaining users:
\begin{equation}
\label{weight1}
\tilde{w}^*=\arg\min _{w}\frac{1}{I-I_{leave}}\sum_{i\in \mathcal{I}\backslash\mathcal{I}_{leave}} F_{i}(w),
\end{equation}
where   $\mathcal{I}_{leave}$ is the set of users who leave the system through federated unlearning. This method is typically more efficient than training from scratch, as the minimum point may not change much after some users leave.

\subsection{Model Accuracy and Unlearning Time}
\label{bounds}
Given the objectives of federated learning and unlearning, we analyze the model accuracy gap and unlearning time in the following.

We use two widely adopted algorithms, Scaffold\cite{karimireddy2020scaffold} and FedAvg \cite{mcmahan2016communication}, as the federated learning algorithms when deriving the optimality gap of the global model. In each local iteration of the algorithm, every user computes a mini-batch gradient with batch size $s_i$. A batch or minibatch refers to equally sized subsets of the training dataset over which the gradient is calculated. In this paper, we consider the widely adopted setting that  users' batch sizes $\{s_i\}_{i\in \mathcal{I}}$ are in the same proportion to  their data sizes $\{d_i\}_{i\in \mathcal{I}}$ (i.e., $s_i=\iota d_i, \forall  i
\in \mathcal{I}, \iota\in (0,1)$)  \cite{bourtoule2021machine,tran2019federated,ding2020optimal}.

The following proposition presents   bounds on the optimality gap for the global models trained with Scaffold or FedAvg:
\begin{proposition}
	\label{modelacc}
	Suppose each user's loss function $F_i$ is $\mu$-strongly-convex and $L$-Lipschitz-smooth. Consider federated learning algorithms  Scaffold and FedAvg with local iteration number of user $i$ denoted by $K_i$ and local step size denoted by $\eta_i$. Set $\bar{\eta} = \eta_i K_i$. Then, we have for Scaffold with $\bar{\eta} \le \frac{1}{12L}$,
	\begin{equation}
	\mathbb{E} \Vert w_{t+1} - w^* \Vert^2 \le (1- \frac{\mu \bar{\eta}}{2}) \mathbb{E} \Vert w_t - w^* \Vert^2 +   \frac{22\bar{\eta}^2\sigma^2}{I}\sum_{i\in \mathcal{I}} \frac{1}{s_i},
	\end{equation}
	where  $w_{t+1}$ and $w_t$ represent the model parameter after global round $t+1$ and $t$, respectively,   $s_i$ is user $i$'s local batch size, and $\sigma^2$ is the variance bound of each data sample.\footnote{To estimate the true gradient $\nabla F_i(w)$, we uniformly sample one data point to generate a gradient estimate  $g_i(\hspace{-0.5mm}w)$  and assume $\mathbb{E} \Vert g_i(\hspace{-0.5mm}w)\hspace{-0.5mm} - \hspace{-0.5mm} \nabla F_i(\hspace{-0.5mm}w) \Vert^2 \hspace{-0.5mm}\le\hspace{-0.5mm} \sigma^2$ for any $w$.} 	
	For FedAvg with $\bar{\eta} \le \frac{1}{12LB}$,
	\begin{equation}
	\begin{split}
	\mathbb{E}\Vert& w^{t+1} - w^* \Vert^2 \le\\& (1 - \frac{\mu \bar{\eta}}{2})\mathbb{E} \Vert w_t - w^* \Vert^2 + 6\bar{\eta}^2 G^2 + \frac{19\bar{\eta}^2\sigma^2}{I}\sum_{i\in \mathcal{I}} \frac{1}{s_i},
	\end{split}
	\end{equation}
	when the bounded dissimilarity assumption is satisfied, i.e., there exist some constants $G \ge 0$ and $B \ge 1$ such that $(1/I)\sum_{i=1}^I \Vert  \nabla F_i(w) \Vert^2 \le G^2 + B^2 \Vert \nabla F(w) \Vert^2$, $\forall w$.
	
	Moreover, by selecting $\bar{\eta} = \frac{c}{t + 1}$ for some $c>0$, we have that the expected optimality gap of the global model satisfies:  for Scaffold with $c \le \frac{1}{12L}$,
	\begin{equation}
	\label{6acc}
	\mathbb{E} \Vert w_t - w^* \Vert^2 \le \frac{1}{t+1}\left( \frac{b_1(c)\sigma^2}{I}\sum_{i\in \mathcal{I}} \frac{1}{s_i} + \Vert w_0 - w^*  \Vert^2\right),
	\end{equation}
	and for FedAvg with $c \le \frac{1}{12LB}$,
	\begin{equation}
	\label{6acc-fedavg}
	\begin{split}
	\mathbb{E} \Vert &w_t - w^* \Vert^2 \le\\& \frac{1}{t+1}\left( \frac{b_2(c)\sigma^2}{I}\sum_{i\in \mathcal{I}} \frac{1}{s_i} + b_2(c) G^2 + \Vert w_0 - w^*  \Vert^2\right),
	\end{split}
	\end{equation}
	where   $b_i(c), i=1,2$ are some monotonically increasing  functions of $c$.
\end{proposition}
The proof of Proposition \ref{modelacc} is given in Appendix A in the technical report \cite{techtmc}. As a large optimality gap $\Vert w_t - w^* \Vert^2$ means a high accuracy loss of the global model,  Proposition \ref{modelacc} presents a relationship between the expected global model accuracy loss and the users' data sizes.  As shown in \eqref{6acc} and \eqref{6acc-fedavg}, the expected accuracy loss of the global model decreases in the users' training batch sizes $\{s_i\}_{i\in\mathcal{I}}$ (and thus data sizes $\{d_i\}_{i\in\mathcal{I}}$).  
Moreover, we explain two asymptotic cases of \eqref{6acc} and \eqref{6acc-fedavg} for better understanding. When the initial point is optimal (i.e., $w_0=w^*$), the bound does not go to zero due to sample randomness. When batch size $s_i$ is large enough,  the randomness is then highly reduced and the bound is controlled by the initialization of the algorithm, i.e., the farther the initial point $w_0$ is from the optimal solution $w^*$, the more iterations are needed. 

Then, after applying  the result in Proposition \ref{modelacc} to the natural unlearning model introduced in Section \ref{unlearningo}, we have the following proposition about federated unlearning rounds:
\begin{proposition}
	\label{unlearntm}
	Consider the same conditions of Proposition \ref{modelacc} with diminishing step size $\bar{\eta}$ and suppose
	$$ 
 b_i(c)\hspace{-0.5mm}\le \hspace{-0.5mm}\frac{1}{(I - I_{leave}) \mu^2} \frac{\left(\sum_{i \in \mathcal{I}_{leave}} \Vert \nabla F_i(w^*) \Vert \right)^2 }{G^2 + \sum_{i \in \mathcal{I}\backslash\mathcal{I}_{leave}}\frac{1}{s_i}\sigma^2},$$ where $i=1$ when using Scaffold and $i=2$ for FedAvg.
It will require
	\begin{equation}
	\label{7ron}
	T_{unlearn} \ge \frac{2(I-1)}{\epsilon^2\mu^2} \sum_{i \in \mathcal{I}_{leave}} \Vert \nabla F_i(w^*) \Vert ^2 - 1
	\end{equation}
rounds of communication to guarantee $\mathbb{E}\Vert w_{T_{unlearn}} - \tilde{w}^* \Vert \le \epsilon$ when  starting from the original learned model $w^*$, where the new model $\tilde{w}^* $ is defined in \eqref{weight1}.
\end{proposition}
The proof of Proposition \ref{unlearntm} is given in Appendix B in the technical report \cite{techtmc}. Each user's gradient  $\Vert \nabla F_i(w^*) \Vert $ can represent his training loss (denoted as $\ell_i$) because the calculated gradient increases in the loss. 
Hence, Proposition \ref{unlearntm} reveals the relationship between the number of communication rounds required for federated unlearning and the training losses of leaving users. 
As indicated in \eqref{7ron}, a larger total training loss of the leaving users  $ \sum_{i \in \mathcal{I}_{leave}} \ell_i ^2 $ (i.e., a larger   $ \sum_{i \in \mathcal{I}_{leave}} \Vert \nabla F_i(w^*) \Vert  ^2 $) requires more communication rounds  $	T_{unlearn}$ to achieve  unlearning.

We will apply the derived results about model accuracy loss and unlearning rounds in building the system model in the next section. 

\section{System Model}
\label{system}
\begin{figure}[tbp]
	\vspace{-0mm}
	\centering
	\includegraphics[width=0.99\linewidth]{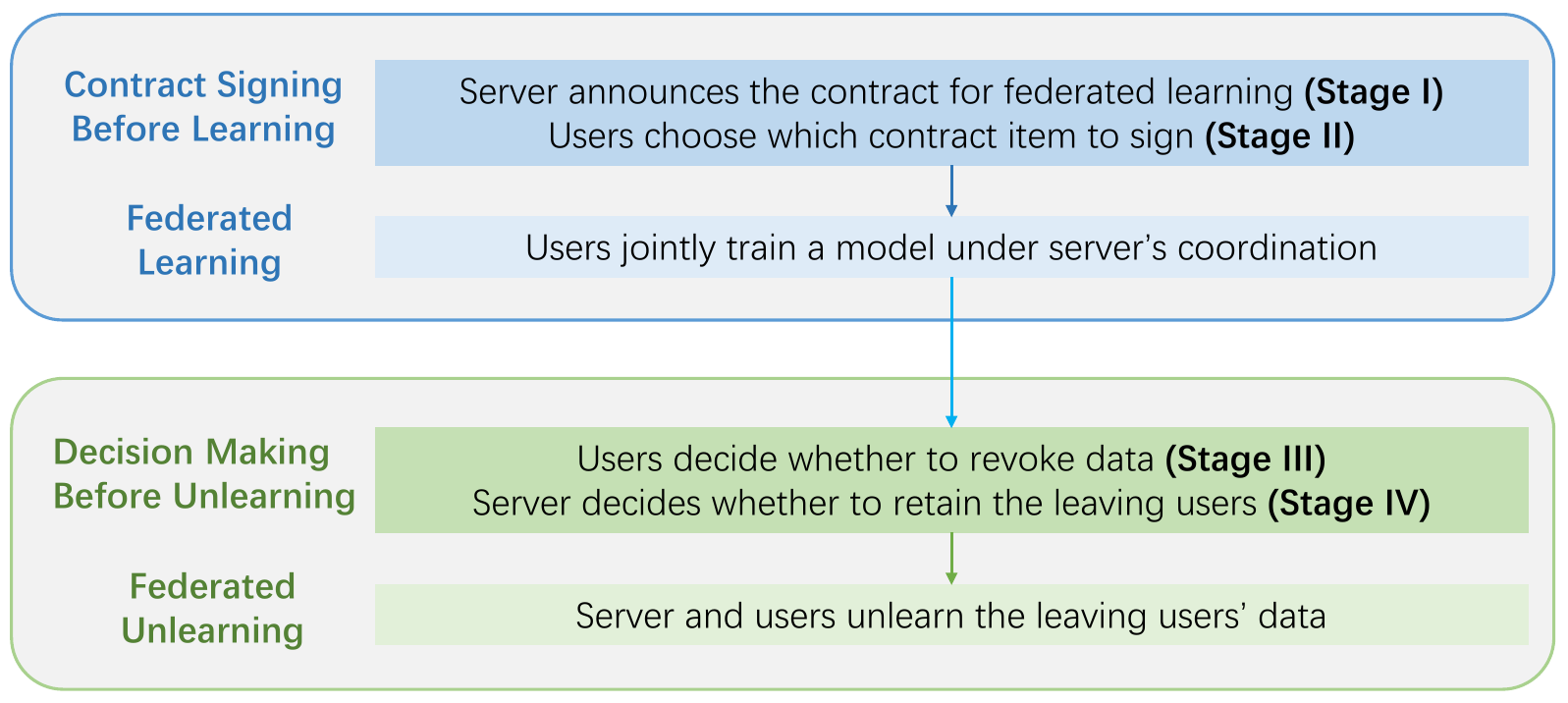}
	\vspace{-3mm}
	\caption{Framework of federated learning and unlearning system with incentive mechanisms.}
	\vspace{-4mm}
	\label{framework}
\end{figure}

We consider a federated learning and unlearning system consisting of a set of heterogeneous users with private data and a central server. As illustrated in Fig.~\ref{framework}, the server first incentivizes users as workers to participate in a federated learning phase through a contract. However, some users may later choose to revoke their data and leave the system. In response,   the server can provide further incentives to retain valuable users. Upon the final exit of some users from the system, the remaining users collectively execute an algorithm to unlearn the leaving users' data.

In the following, we first divide the heterogeneous users into different types for the convenience of incentive design,  then formulate a multi-stage game between the strategic server and users, and finally specify the payoffs of the server and the users (i.e., their optimization objectives) in two unlearning scenarios, respectively.

\subsection{User Type}
We consider a set $\mathcal{I}\triangleq\{1,2,..., I\}$ of users in the system with two-dimensional private information: marginal cost for training effort  $\theta$ and marginal perceived privacy cost $\xi$. We refer to a user with $(\theta_j,\xi_j)$ as a type $j$ user. We further assume that the $I$ users belong to a set $\mathcal{J}\triangleq\{1,2,...,J\}$ of $J$ types. Each type $j$ has $I_j$ users, with $\sum_{j \in \mathcal{J}} I_{j}=I$. 
The total number of users $I$ and the number of each type $I_j$ are public information, but each user's specific type is private information.\footnote{The server can have knowledge about statistics of type information through market research and past experiences, but it is hard for it to know each user's private type.}

Under private information, it is difficult for the server to predict users' strategies. To this end, we propose to design a contract mechanism for the server to elicit information.

\subsection{Games and Strategies}
We use a multi-stage Stackelberg game to model the interaction between the server and users in
each of the two scenarios.
\subsubsection{Unlearning-Allowed Scenario}
When unlearning is allowed, we consider  the following four-stage game that captures the move sequence of the server and the users: 
\begin{itemize}
	\item Stage I:  The server designs a  federated learning incentive contract $\boldsymbol{\phi}\triangleq \left\{\phi_{j}\right\}_{j \in \mathcal{J}}$, which contains $J$ contract items (one for each user type). Each contract item $\phi_{j} \triangleq\left(d_{j}, r^L_{j}\right)$ specifies the relationship between the required data size $d_j$  of each type-$j$ user (for local computation) and the corresponding learning reward $r^L_j$. 

	\item Stage II: Users decide which contract item to choose. Then,  they jointly implement the federated learning algorithm (Scaffold or FedAvg).

	\item Stage III: 
	Users decide whether to revoke data after federated learning. We denote a user $i$'s revocation decision as 
	\begin{equation}
	\begin{split}
	x_i=  \left\{ 
	\begin{array}{rcl}
	& 0,  &\rm{if }\; \text{user $i$ does not revoke data},\\
	& 1,    &\rm{if }\; \text{user $i$ revokes his data},\\
	\end{array} \right.
	\end{split}
	\end{equation}
	and denote the set of users who revoke their data as $\mathcal{I}_u$. If a type-$j$ user revokes his data, then he needs to fully return the reward $r_j^L$ to the server.\footnote{If there is no such return policy, every user can first participate to get rewards and then revoke data to reduce costs, resulting in a catastrophic failure of model training collaboration and a huge cost to the server.} We consider that the server will announce users' training losses $\{\ell_i\}_{i\in \mathcal{I}}$  (without specifying users)  after federated learning to help users decide whether to revoke data.\footnote{It is not obvious that a strategic server would make such an announcement, but it can be stipulated by regulations for protecting users' right to be forgotten. If we do not make this assumption, the problem will be even simpler. As we shall see in the analysis in Section \ref{siii}, we just need to replace other users' training losses $\{\ell_k\}_{k\in \mathcal{I}}$ in \eqref{22} with the same expected loss $\mathbb{E}[\ell]$ and solve the problem through a similar approach.} 
	\item Stage IV: The server decides the set of leaving users to retain $\mathcal{I}_r$ and designs the corresponding retention incentives $\left\{r^U_{i}\right\}_{i \in \mathcal{I}_r}$, 
	such that those receiving the retention incentives will choose to stay in the system and those without will leave.\footnote{In this case,   $\mathcal{I}_u\backslash\mathcal{I}_r$ is the set of users who finally leave the system, and $\mathcal{I}\backslash(\mathcal{I}_u\backslash\mathcal{I}_r)$ is the set of users who finally stay.} The remaining users and server collectively implement federated unlearning.
\end{itemize}

\begin{table}[tbp]
	\vspace{-0mm}
	\caption{The Server and Users' Knowledge in Different Stages}  
	\vspace{-4mm}
	\begin{center}
		\begin{tabular}{|c|c|c|}
			\hline
			\textbf{Stage}&  \textbf{Known} & \textbf{Unknown} \\
			\hline
			\tabincell{c}{Server in Stage I} &  \tabincell{c}{$\mathcal{J}$,  $\{I_j\}_{j\in \mathcal{J}}$} &  \tabincell{c}{$\{\theta_i,\xi_i,\ell_i,v_i\}_{i\in \mathcal{I}}$ }\\
			\hline
			\tabincell{c}{User in Stage II} & \tabincell{c}{his own type $(\theta_i,\xi_i)$ }& \tabincell{c}{other users' types,\\ $\{\ell_i,v_i\}_{i\in \mathcal{I}}$ }\\
			\hline
			\tabincell{c}{User in Stage III} &  \tabincell{c}{his own type $(\theta_i,\xi_i)$, \\$\{\ell_i\}_{i\in \mathcal{I}}$ } &\tabincell{c}{other users' types, \\$\{v_i\}_{i\in \mathcal{I}}$}\\
			\hline
			\tabincell{c}{Server in Stage IV} & \tabincell{c}{  $\mathcal{J}$,  $\{I_j\}_{j\in \mathcal{J}}$,\\ $\{\theta_i,\xi_i,\ell_i,v_i\}_{i\in \mathcal{I}}$} &\diagbox{ }{ }\\
			\hline
		\end{tabular}
		\label{sce} 
		\vspace{-4mm}
	\end{center}
\end{table}

\begin{table}[tbp]
	\caption{Key Notations}  
	\vspace{-4mm}
	\begin{center}
		\begin{tabular}{|c|c|}
			\hline
			$\theta_j$ &  Marginal training cost of type-$j$ users \\
			\hline
			$\xi_j$ & Marginal perceived privacy cost of type-$j$ users \\
			\hline
			$I_j$ & Number of type-$j$ users \\
			\hline
			$j/\mathcal{J}$ &  Index/Set of user types in the system \\
			\hline
			$i/\mathcal{I}$ &  Index/Set of users in the system \\
			\hline
			$\mathcal{I}_u$ &  Set of users who revoke their data in Stage III \\
			\hline
			$\mathcal{I}_r$ &  Set of users who are retained by the server in Stage IV \\
			\hline
			$\phi_{j}$ &   Contract item designed for type-$j$ users \\
			\hline
			$d_j$ &   Required  data size for each type-$j$ user in the contract\\
			\hline
			$r_j^L$ & Learning reward for each type-$j$ user in the contract\\			
			\hline
			$r_i^U$ & Unlearning reward (retention incentive) for   user $i$\\
			\hline		
			$x_i$& User $i$'s data revocation decision\\
			\hline
			$p_j$ & Historical revocation rate of type-$j$ users\\
			\hline
			$q_j$ & Historical retention rate of type-$j$ users\\
			\hline
			$T$ &  Number of communication rounds of federated learning \\
			\hline
			$\lambda$ & Coefficient related to unlearning communication rounds\\
			\hline
			$\varrho$& Coefficient related to expected accuracy loss\\
			\hline
			$\gamma$ & Server's weight on incentive rewards\\		
			\hline
			$v_i$ & User $i$'s contribution to global model accuracy\\
			\hline
			$\ell_i$ & User $i$'s training loss (representing $||\nabla F_i(x^*)||$) \\
			\hline
		\end{tabular}
		\label{tablit} 
		\vspace{-5mm}
	\end{center}
\end{table}

In Stage III, we use $\ell_i=||\nabla F_i (w_T)||$ to represent the training loss, where $w_T$ is the solution obtained after $T$ iterations of Scaffold or FedAvg. We assume $T$ is large enough, such that $w_T$ and $w^*$ are close. A large  $\ell_i$ implies the federated solution is far away from the minimizer of local loss function $F_i$ and therefore a larger training loss. 

After federated learning, the server and users have more information in Stages III and IV compared with Stages I and II. For example, the users will know their training losses $\{\ell_i\}_{i\in \mathcal{I}}$.  The server can evaluate the users' contribution to the global model (denoted by $\{v_i\}_{i\in \mathcal{I}}$), and it will know each user's type by observing users' contract item choices.  We summarize their knowledge about some key information in the four stages in Table \ref{sce} and  list the key notations  in this paper in Table \ref{tablit}.\footnote{As analyzing the four-stage game is complicated, this paper does not model the information update in a fully Bayesian framework but specifies plausible beliefs that the players hold in each stage.}

Moreover, in Stage IV, the server has enough information to know whether the users will accept the retention incentives. Therefore, we do not model a Stage V   in which the users decide to accept or not accept the retention incentives. 
After that, as in Fig.~\ref{framework}, the staying users perform federated unlearning under the server's coordination, which makes staying users sustain unlearning costs. We will specify the payoffs and costs of the server and users in each stage of the game in the next subsection.

\subsubsection{Unlearning-Forbidden Scenario}
Without the unlearning process, we consider a two-stage game that only includes Stages I and II from the unlearning-allowed case.

\subsection{Payoffs  in the Unlearning-Allowed Scenario}
\label{payoffall}
At each stage,  every user or the server seeks to maximize his expected payoff (or minimize his expected cost) based on his current knowledge. As knowledge updates occur between stages, the payoffs of the users or the server (maximization or minimization objectives respectively) take different forms in each stage.

\subsubsection{Server's Payoff in Stage I}
The server's objective in Stage I is to minimize the sum of the expected accuracy loss of the global model and the expected total incentive rewards for users.

First, we specify the expected model accuracy loss, which depends on the data of users who finally stay in the system. 
Since the server cannot predict which users will leave and who to retain due to the lack of information in Stage I, it can only base its decision on user distribution expectations. 
Specifically, we assume that according to the historical experience and market statistics, the server knows the probability of a type-$j$ user revoking his data (i.e., his revocation rate) $p_j$  and the probability that a type-$j$ user who wants to revoke data is retained  (i.e., his retention rate) $q_j$, where $p_j$ and $q_j$ are independent. 
Following Proposition \ref{modelacc}, we   model the server's expected accuracy loss after federated unlearning as:
\begin{equation}
\frac{\varrho}{T}\sum_{j \in \mathcal{J}}I_j(1-p_j+p_jq_j)\frac{1}{d_j },
\end{equation}
where $T$ is the number of communication rounds of federated learning, $\varrho$ is a coefficient related to the sample variance, and $1- p_j +p_jq_j$ is the percentage of type $j$ users remaining in the system in the end. This captures that the expected model accuracy loss decreases in the data sizes of all staying users.\footnote{As the server aims to incentivize users to contribute data in federated learning, we only model the impact of data sizes and omit the independent term about initial point $w_0$ in \eqref{6acc} and \eqref{6acc-fedavg}. Since   we consider that  users' batch sizes $\{s_i\}_{i\in \mathcal{I}}$ are in the same proportion to  their data sizes $\{d_i\}_{i\in \mathcal{I}}$,  it is equivalent to  substitute $s_i$ with $d_i$ in \eqref{6acc} and \eqref{6acc-fedavg}.}

The server's payoff also includes the cost of all rewards it pays to users, which comprises the initial contract announced in Stage I and incentives offered to encourage leaving users to remain in Stage IV. 
If all users choose to participate in the contract and choose their corresponding contract items,\footnote{As we shall see in Section \ref{si},  we will design	the contract to ensure that each user will participate  (i.e., individual rationality) and choose the contract item designed	for his type (i.e., incentive compatibility).} the expected total learning reward is $\sum_{j \in \mathcal{J}}I_j(1-p_j+p_jq_j)r^{L}_j$. Note that if a type-$j$ user successfully revokes his data, he needs to fully return the reward $r^{L}_j$ to the server. 
The server's expected incentive for retaining leaving users is $\mathbb{E}[\sum_{i \in \mathcal{I}_r}r_i^U]$, which depends on $p$, $q$, and training losses and will be calculated through backward induction in Section \ref{si}.

Combining these terms, the server's expected cost in Stage I is 
\begin{equation}
\label{s1}
\begin{split}
W^{s-1}&=\frac{\varrho}{T}\sum_{j \in \mathcal{J}}I_j(1-p_j+p_jq_j)\frac{1}{d_j}\\&+\gamma\left(\sum_{j \in \mathcal{J}}I_j(1-p_j+p_jq_j)r^L_j+\mathbb{E}\bigg[\sum_{i \in \mathcal{I}_r}r_i^U\bigg]\right),
\end{split}
\end{equation}
where $\gamma$  is how much weight the server puts on the incentive reward payments compared to the model accuracy loss. A smaller $\gamma$ means that the server is less concerned about minimizing the incentive rewards and more concerned about reducing the accuracy loss.

\subsubsection{Users' Payoffs in Stage II}
In the overall game, there are three possible outcomes for a user (not revoke data, revoke and retained, revoke and not retained). However, in this stage, a user does not have enough information to know which outcome will realize, so he must calculate his expected payoff by considering three cases:
\begin{itemize}
	\item \emph{Case (a): not revoke.} With probability $1-p_j$, a type-$j$ user will not revoke his data after federated learning. In this case, his expected payoff is the difference between the learning reward $r^L_j$ and  costs (including the learning cost, privacy cost, and unlearning cost):
	\begin{equation}
	\label{notr}
	U_{j,a}^{s-2}=r^L_j-\theta_jd_jT-\xi_j\mathbb{E}[\ell_j]d_j-\mathbb{E}\bigg[\theta_jd_j\lambda\sum_{i \in \mathcal{I}_u\backslash\mathcal{I}_r}\ell_i^2\bigg],
	\end{equation}
where $\theta_jd_jT$ is  the total learning cost in $T$ rounds. As we consider that each user's sampled data size in each local round is proportional to his total data size,  the learning cost is linear in his data size $d_j$ 
	(e.g., \cite{bourtoule2021machine,tran2019federated,ding2020optimal}).   Similarly, in the unlearning cost $\theta_jd_j\lambda\sum_{i \in \mathcal{I}_u\backslash\mathcal{I}_r}\ell_i^2$, the $\lambda\sum_{i \in \mathcal{I}_u\backslash\mathcal{I}_r}\ell_i^2$ models the number of communication rounds for unlearning, which increases in the leaving users' training losses (according to Proposition \ref{unlearntm}).\footnote{We use the simplified model of \eqref{7ron} in Proposition \ref{unlearntm} to capture the key relationship between the unlearning communication rounds $T_{unlearn}$ and leaving users' training losses (represented by   $\Vert \nabla F_i(w^*) \Vert $).} 
	A type $j$ user's perceived privacy cost  $\xi_j\mathbb{E}[\ell_j]d_j$ increases in his expected training loss $\mathbb{E}[\ell_j]$   and data size $d_j$. As a high training loss  $\ell_j$ reflects a large distance of user $j$'s data from the average of other users' distribution, we use it to measure the uniqueness of a user. Thus, the model captures that the privacy cost increases in the uniqueness and size of one's training data (e.g., \cite{de2013unique,romanini2021privacy}). As each user cannot know his exact training loss $\ell_j$ before federated learning, we assume that he estimates the expected loss using the public distribution (with mean $\mathbb{E}[\ell_j]$ and variance $D(\ell_j)$). 
	\item \emph{Case (b): revoke but retained.} With probability $p_jq_j$, a type-$j$ user will revoke his data after federated learning but will be retained by the server through more incentives $r^U_j$. In this case, his expected payoff is the difference between total rewards (including both learning and unlearning incentives) and costs:
	\begin{equation}
	\label{rere}
	\begin{split}
	U_{j,b}^{s-2}=&r^L_j+\mathbb{E}\left[r^U_j\right]-\theta_jd_jT\\&-\xi_j\mathbb{E}[\ell_j]d_j-\mathbb{E}\bigg[\theta_jd_j\lambda\sum_{i \in \mathcal{I}_u\backslash\mathcal{I}_r}\ell_i^2\bigg].
	\end{split}
	\end{equation}
	The unlearning incentive $r^U_j$ will be determined by the server in Stage IV based on users' training losses, contributions, and data revocation, which are unknown in this stage. Thus, each user can only calculate the expectation of the unlearning incentive.
	\item \emph{Case (c): revoke and not retained.} With probability $p_j(1-q_j)$, a type-$j$ user will revoke his data and will not be retained by the server, i.e.,  the user's data will be unlearned.  The user needs to return the reward $r^L_j$ to the server but will not incur any privacy cost or unlearning cost. In this case, his expected payoff is
	\begin{equation}
	\label{renre}
	U_{j,c}^{s-2}=-\theta_jd_jT,
	\end{equation}
	which is the sunk training cost from federated learning.
\end{itemize}
In summary, a type-$j$ user's expected payoff  in Stage II is 
\begin{equation}
\label{10}
\begin{split}
U_{j}^{s-2}=(1-p_j)U_{j,a}^{s-2}+p_jq_jU_{j,b}^{s-2}+p_j(1-q_j)U_{j,c}^{s-2}.\\
\end{split}
\end{equation}
%
%
%
If  $U_j^{s-2}\ge 0$, the type-$j$ user will choose to participate in the federated learning in Stage II.

\subsubsection{Users' Payoffs in Stage III}
After federated learning, each user $i$ has knowledge about his training loss $\ell_i$. If   user $i$ chooses not to revoke his data, his expected payoff in Stage III is (updating \eqref{notr} in Case (a) with the realized training loss $\ell_i$): 
\begin{equation}
U_{i,a}^{s-3}=r^L_i-\theta_id_iT-\xi_i\ell_id_i-\mathbb{E}\bigg[\theta_id_i\lambda\sum_{k \in \mathcal{I}_u\backslash\mathcal{I}_r}\ell_k^2\bigg].
\end{equation}
The reason for using expectation here is that users do not know the set of retained users $\mathcal{I}_r$   determined in Stage IV.
Users' expected payoffs of Cases (b) and (c) in Stage III follow the same approach (i.e., updating \eqref{rere} and \eqref{renre} with the realized training loss $\ell_i$).
%
%



Note that users of the same type may have different training losses and thus different payoffs, so the payoff in Stage III is user-specific instead of type-specific. 
Moreover, after some users leave, the remaining users' training losses may change as the global model will be updated. Since users cannot accurately predict their future expected loss even if they know all users' current losses, we assume that each user still approximates his future expected loss as equal to his current loss.


\subsubsection{Server's Payoff in Stage IV}
\label{piv}
When some users want to leave the system, it is important for the server to know their contributions to the global model for retaining valuable users.

A fair and effective method to compute a user's contribution to a coalition is the Shapley value \cite{winter2002shapley}.  Wang et al. \cite{wang2020principled} introduced a related concept called federated Shapley value to evaluate each user's contribution in a federated learning setting. The federated Shapley value for user $i$, denoted as $v_i$, is calculated by the server during the federated learning process and is unknown to the users.


Once obtaining users' contributions (federated Shapley values), the server can calculate its realized cost in Stage IV. This cost is the sum of two factors: the realized accuracy loss, which is estimated by the sum of federated Shapley values of all users who remain in the system, and the realized incentives.
\begin{equation}
\label{19}
W^{s-4}=\sum_{i \in \mathcal{I}\backslash(\mathcal{I}_u\backslash\mathcal{I}_r)}v_i+\gamma\left(\sum_{i \in \mathcal{I}\backslash(\mathcal{I}_u\backslash\mathcal{I}_r)}r^L_i+\sum_{i \in \mathcal{I}_r}r^U_i\right).
\end{equation}
The first term in \eqref{19} represents the model accuracy loss, the second is the learning reward paid to all remaining users for participation in federated learning, and the last term is the total retention incentive. The additivity property of federated Shapley values allows the server to compare all the possible sets of users to retain and find the optimal one. Note that a smaller federated Shapley value is better, as it means a larger contribution to the accuracy of the global model, and the federated Shapley values can be negative.


\subsection{Payoffs in the  Unlearning-Forbidden Scenario}
Similar to Stages I and II in the unlearning-allowed scenario, we now specify the users and the server's payoffs in the unlearning-forbidden scenario. The difference here is that there are no unlearning considerations (e.g., data revocation or retention incentives). We will use the superscript $^\prime$ for the unlearning-forbidden scenario to differentiate the notations in the two scenarios. 
\subsubsection{Server's Payoff in Stage I}
The server needs to minimize the sum of the expected accuracy loss and the  incentive rewards paid for federated learning:
\begin{equation}
\label{uw}
W=\frac{\varrho}{T}\sum_{j' \in \mathcal{J}'}\frac{I'_{j'}}{d'_{j'}}+\gamma\sum_{j' \in \mathcal{J}'}I'_{j'}{{r}_{j'}^{L\prime}}.
\end{equation}

\subsubsection{Users' Payoffs in Stage II}
A type-$j'$ user's payoff is the difference between the learning reward and training costs (including the learning cost and perceived privacy cost)
\begin{equation}
U_{j'}=r^{L\prime}_{j'}-\theta_{j'}d'_{j'}T-\xi'_{j'}\mathbb{E}[\ell'_{j'}]d'_{j'}.
\end{equation}
Note that the marginal perceived privacy cost $\xi'_{j'}$ here should be no smaller than that in the unlearning-allowed scenario $\xi_{j}$ for the same user, as a user cannot revoke his data once he decides to participate in the federated learning in the unlearning-forbidden scenario.

Next, we will use the standard backward induction to analyze the server and users’ optimal
strategies in two unlearning scenarios.

\section{Optimal Incentive Mechanism in Unlearning-Allowed Scenario}
\label{allowed}
In this section, we analyze an optimal incentive mechanism for the unlearning-allowed scenario. Based on backward induction, we will derive the optimal strategies from Stage IV to Stage I in Sections \ref{siv}-\ref{si}, respectively. 
\subsection{Server's Retention Strategies in Stage IV}
\label{siv}
Given the server's contract $\boldsymbol{\phi}$ in Stage I, the users' contract item choices in Stage II, and the users' revocation decisions $\mathcal{I}_u$ in Stage III, the server needs to determine which users to retain $\mathcal{I}_r$ and the corresponding retention incentives $\{r^U_i\}_{i\in \mathcal{I}_r}$ in Stage IV. 

As we discussed  in Section \ref{piv},  the server seeks to minimize the cost in \eqref{19} in Stage IV, which can be formulated as follows: 
\begin{problem}[Server's Optimization Problem in Stage IV]
	\label{p1}
	\begin{subequations}
		\begin{align}
		\min \;&  \sum_{i \in \mathcal{I}\backslash(\mathcal{I}_u\backslash\mathcal{I}_r)}v_i+\gamma\left(\sum_{i \in \mathcal{I}\backslash(\mathcal{I}_u\backslash\mathcal{I}_r)}r^L_i+\sum_{i \in \mathcal{I}_r}r^U_i\right)\label{16a}\\
		\rm{s.t.}\; &r^U_i  +r^L_i-\theta_id_iT-\xi_i\ell_id_i-\theta_id_i\lambda\sum_{k \in \mathcal{I}_u\backslash\mathcal{I}_r}\ell_k^2\notag\\&\hspace{37mm}\ge -\theta_id_iT, \forall  i \in  \mathcal{I}_r\label{16b}	\\
		\rm{var.}\; &\mathcal{I}_r\subseteq\mathcal{I}_u, \{r^U_i\}_{i \in  \mathcal{I}_r}.\label{16c}
		\end{align}
	\end{subequations}
\end{problem}
The constraint \eqref{16b} is to ensure that the retention incentives are enough to make the target users stay in the system. The left-hand side of the constraint is a user $i$'s payoff after accepting the retention incentive (including unlearning reward, learning reward, learning cost, privacy cost, and unlearning cost), and the right-hand side is his payoff of not accepting (i.e.,  he has to return the learning reward to the server and only has sunk learning cost). 

The following proposition presents the solution to Problem \ref{p1}. 
\begin{proposition}
	\label{lm}
	The  server's optimal set of users to retain is
	\begin{equation}
	\label{20}
	\mathcal{I}_r^*=\arg\min_{\mathcal{I}_r\subseteq\mathcal{I}_u} \; \sum_{i \in \mathcal{I}_r}\left(v_i+\gamma \theta_id_i\lambda\sum_{k \in \mathcal{I}_u\backslash\mathcal{I}_r}\ell_k^2+\gamma\xi_i\ell_id_i\right),
	\end{equation}
	and the optimal retention incentives are
	\begin{equation}
	\label{21}
	{r^{U}_i}^*=\theta_id_i\lambda\sum_{k \in \mathcal{I}_u\backslash\mathcal{I}_r^*}\ell_k^2+\xi_i\ell_id_i-r^L_i, \forall  i \in  \mathcal{I}_r^*.
	\end{equation}
\end{proposition}
The proof of Proposition \ref{lm} is given in Appendix C in the technical report \cite{techtmc}. 
Proposition \ref{lm} highlights a trade-off regarding the retention of users and their training losses. Users who have larger training losses incur higher privacy costs and thus require higher incentives to retain (indicated by $\gamma\xi_i\ell_id_i$ in \eqref{20}). However, retaining such users also helps reduce the unlearning costs since the objective in \eqref{20} increases with the aggregated loss of the leaving users. 
Furthermore, the server has the incentive to retain users who contribute more to the model accuracy, which corresponds to smaller values of $v_i$. Additionally, users with smaller marginal costs  $\theta_i$ and $\xi_i$ are also desirable to reduce unlearning incentives.\footnote{Note that in \eqref{20}, the server may not only include users with a negative value in the brackets, as retaining some users with positive values may reduce the server's objective through the aggregated losses. This is an integer programming problem with complexity $\mathcal{O}(2^{I_u})$. When the number of leaving users $I_u$ is large, the server can reduce the complexity by classifying the leaving users into several categories to retain, each category with similar contributions and costs.}

\subsection{Users' Revocation Decisions in Stage III}
\label{siii}
Considering the server's optimal retention strategies in Stage IV, each user $i$ decides whether to revoke his data in Stage III given the information announced in Stages I and II.


Based on the server's optimal retention incentives \eqref{21} and the user's payoffs in Stage III (i.e., the updated \eqref{rere} and \eqref{renre} with realized losses),   a user $i$'s payoff after revoking data is $-\theta_id_iT$, regardless of whether the user is retained by the server or not. 
Thus,  user $i$'s expected payoff in Stage III can be rewritten as
\begin{equation}
\label{22}
\begin{split}
&U_i^{s-3}(x_i;x_{-i})=x_i\left(-\theta_id_iT\right)\\&+(1-x_i)\left[r^L_i-\theta_id_iT-\xi_i\ell_id_i-\theta_id_i\lambda\sum_{k \in \mathcal{I}}x_k(1-q)\ell_k^2\right],
\end{split}
\end{equation}
where $x_{-i}=\{x_k\}_{k\in \mathcal{I}\backslash\{i\}}$ is the revocation decisions of all users except user $i$ and   $q=\mathbb{E}[q_j]$ is the expected retention rate of all users, as users do not know each other's type.\footnote{Here we use the historical retention rate $q$  to calculate the expected payoffs instead of the retention rate obtained in Stage IV (i.e., ${|\mathcal{I}_r^*|}/{|\mathcal{I}_u|}$). This is because users do not know their federated Shapley values and cannot calculate $\mathcal{I}_r^*$. If they calculate the expectation $\mathbb{E}[\mathcal{I}_r^*]$ based on type statistics, according to \eqref{20}, the result will be user type retention instead of user retention (e.g., retain all type-i users and not retain all type-$j$ users regardless of different data distributions and losses of the same type of users), which is not true. 
	Conversely, historical rates ranging between $[0,1]$ allow for more
	realistic partial retention of same-type users. 
	Therefore, we assume that the users have a belief at this stage in the retention rate which is the same as the historical rate.  In the following analysis in Stages I and II, we will also use the historical rates for calculating the expected cost/payoffs for similar reasons. 
} 
As shown in \eqref{22}, each user's payoff depends on the other users' revocation decisions, so users engage in a non-cooperative game in Stage III.  	

We formally define users' non-cooperative sub-game as follows.
\begin{game}[Users' Revocation Sub-Game in Stage III]\quad \\
	\vspace{-4.5mm}
	\begin{itemize}
		\item Players: all users in set $\mathcal{I}$.
		\item Strategy space: each user $i\in \mathcal{I}$ decides whether to revoke his data, i.e., $x_i\in \{0,1\}$ (0: not revoke, 1: revoke).
		\item Payoff function: each user $i\in \mathcal{I}$ maximizes his payoff in \eqref{22}.
	\end{itemize}
	\label{game2}
\end{game}

The following proposition characterizes the Nash equilibrium (NE) of Sub-Game \ref{game2}:
\begin{proposition}
	\label{pro2}
	Sub-Game \ref{game2} is a supermodular game, where pure NE  exists but may not be unique. Algorithm \ref{alg:B} converges to one NE.
\end{proposition}

\setlength{\intextsep}{0pt}
\setlength{\textfloatsep}{0pt}
\begin{algorithm}
	\caption{Users' optimal revocation decisions}
	\label{alg:B}
	\SetKwInOut{Input}{Input}\SetKwInOut{Output}{Output}
	\Input  {$\{r^L_i,\xi_i,\ell_i,d_i,\theta_i\}_{i\in \mathcal{I}},\lambda,q$}
	\Output  {Optimal revocation decisions $\{x_i^*\}_{i\in \mathcal{I}}$}
	
	Initialize $x_i^*\leftarrow 0, i\in \mathcal{I}$;
	
	\While{$\exists x_i^*= 0\; \&\; r^L_i-\xi_i\ell_id_i-\theta_id_i\lambda(1-q)\sum_{k \in \mathcal{I}\backslash\{i\}}x_k\ell_k^2< 0$}{$x_i^*\leftarrow 1,\forall i\; \text{satisfying conditions in line 2}; $}
\end{algorithm}
The proof of Proposition \ref{pro2} is given in Appendix D in the technical report \cite{techtmc}.  Based on Algorithm \ref{alg:B}, we can find the set of users who revoke data in one NE, i.e.,  $\mathcal{I}_u^*=\{i:x_i^*=1,i\in \mathcal{I}\}$. 
Basically, Algorithm \ref{alg:B} corresponds to doing best response updates of the users starting from all users choosing not to revoke (i.e., 0).  It is well known that for supermodular games, these updates will converge monotonically to a NE. 
Algorithm \ref{alg:B} will terminate within
$I$ iterations.\footnote{We can also initialize all the users’ decisions as 1 and check whether there exists a user who wants to change his action 
	from 1 to 0 for payoff improvement. If the equilibrium is the same as that found by Algorithm \ref{alg:B},  it is the unique NE, as Game \ref{game2} is a supermodular game.} The resulting equilibrium strategies and insights will be illustrated through simulation in Section \ref{iii}. 



\subsection{Users' Contract Item Choices in Stage II}
\label{sii}

Based on the analysis in Stages III and IV,  a type-$j$ user's expected payoff in Stage II \eqref{10} can be rewritten as:
\begin{equation}
\begin{split}
U_j^{s-2}=(1-p_j)r_j^L-\kappa_jd_j,
\end{split}
\end{equation}
where 
\begin{equation}
\begin{split}
\kappa_j&\triangleq(1-p_j)\xi_j\mathbb{E}[\ell_j]+\theta_jT\\&+\theta_j (1-p_j)\lambda\sum_{m \in \mathcal{J}}I_mp_m(1-q_m)\left(\mathbb{E}[\ell_m]^2+D(\ell_m)\right),
\end{split}
\end{equation}
 and $D(\ell_m)$ is the variance of type-$m$ users' training losses.

Each type-$j$ user in Stage II will choose a contract item that gives him a maximum non-negative expected payoff, leading to the constraints that the server needs to consider in Stage I.

\subsection{Server's Contract in Stage I}
\label{si}
In Stage I, the server designs a contract to minimize its expected cost, considering the results in Stages II-IV. 

When designing the contract, the server needs to ensure that each user achieves a non-negative payoff, so that the user will accept the corresponding contract item. Moreover, since the server does not know each user's type in Stage I, the server also needs to make a user choose the contract item intended for him (i.e., the user does not misreport his type).\footnote{Revelation principle demonstrates that if a social choice function can be
	implemented by an arbitrary mechanism, then the same function can be implemented by an incentive-compatible-direct-mechanism (i.e. in which users
	truthfully report types) with the same equilibrium outcome. Thus, requiring
	IC will simplify the mechanism design without affecting optimality.}
In other words, a contract is feasible if and only if it satisfies Individual Rationality (IR) and Incentive Compatibility (IC) constraints:
\begin{defn}[Individual Rationality]
	\label{dir}
	A contract is individually rational if   each  type-$j$ user receives a non-negative payoff by accepting the contract item $\phi_j=\left(d_{j}, r^L_{j}\right)$ intended for his  type, i.e.,
	\begin{equation}
	\label{IR}
	(1-p_j)r^L_j-\kappa_jd_j\ge 0, \forall j\in \mathcal{J}.
	\end{equation}
\end{defn}
\begin{defn}[Incentive Compatibility]
	\label{dic}
	A contract is incentive compatible if each type-$j$ user maximizes his own payoff by choosing the contract item $\phi_j=\left(d_{j}, r^L_{j}\right)$ intended for his type, i.e., 
	\begin{equation}
	\label{IC}
	(1-p_j)r^L_j-\kappa_jd_j\ge (1-p_j)r^L_{m}-\kappa_jd_{m}, \forall j, m \in \mathcal{J}.
	\end{equation}
\end{defn}

Considering the constraints in Definitions \ref{dir} and \ref{dic}, the server in Stage I seeks to design the contract $\boldsymbol{\phi}=\{(d_{j}, r^L_{j})\}_{j \in \mathcal{J}}$ to minimize its expected cost in \eqref{s1}, which is rewritten as follows after combining the results in  Stages II-IV: 
\begin{problem}
	\label{pro3}
	\begin{equation}
	\begin{split}
	\min \;& \sum_{j \in \mathcal{J}}\left(\frac{\varrho I_j(1-p_j+p_jq_j)}{Td_j} +\gamma I_j(1-p_j)r^L_j\right.\\&\hspace{28mm}+\gamma I_jp_jq_j(\alpha\theta_j+\xi_j\mathbb{E}[\ell_j])d_j\bigg),\\
	\rm{s.t.}\; &(1-p_j)r^L_j-\kappa_jd_j\ge 0, \forall j\in \mathcal{J},	\\
	&(1-p_j)r^L_j-\kappa_jd_j\ge (1-p_j)r^L_{m}-\kappa_jd_{m}, \forall j, m \in \mathcal{J},	\\
	\rm{var.}\; &\left\{\left(d_{j}, r^L_{j}\right)\right\}_{j \in \mathcal{J}},
	\end{split}
	\end{equation}
	where 
	\begin{equation}
	\alpha\triangleq\lambda\sum_{j \in \mathcal{J}}I_jp_j(1-q_j)\left(\mathbb{E}[\ell_j]^2+D(\ell_j)\right).
	\end{equation}
\end{problem}
Solving Problem \ref{pro3} involves two challenges. First, users' multi-dimensional heterogeneity leads to a challenging multi-dimensional contract design for the server. We will simplify the analysis by summarizing users' multi-dimensional heterogeneity into several one-dimensional metrics,  to guide the server's design of the optimal rewards and data sizes in the contract.
Second, as the total number of IR and IC constraints is large (i.e., $J^2$), it is challenging to obtain the optimal contract directly. To overcome such a complexity issue,  we will first transform the constraints into a smaller number of equivalent ones (Lemma \ref{lm1}).  Then, for any given data size $\boldsymbol{d}=\{d_j\}_{j\in \mathcal{J}}$, we derive the server's optimal reward $\{r^{L*}_{j}(\boldsymbol{d})\}_{j\in \mathcal{J}}$ (Lemma \ref{reward}) in Section \ref{subr}. Finally, we derive the optimal data size $\boldsymbol{d}^*$ (Proposition \ref{prop2} and Theorem \ref{thm1}) in Section \ref{subd}.

\subsubsection{Optimal Rewards in Contract}
\label{subr}
Without loss of generality, we assume that users are indexed in ascending order of 
\begin{equation*}
\begin{split}
\pi_j\triangleq\frac{\kappa_j}{1-p_j}&=\xi_j\mathbb{E}[\ell_j]+\frac{\theta_jT}{1-p_j}\\&+\theta_j \lambda\sum_{m \in \mathcal{J}}I_mp_m(1-q_m)\left(\mathbb{E}[\ell_m]^2+D(\ell_m)\right),
\end{split}
\end{equation*}
which can be regarded as a type-$j$ user's aggregated marginal cost. 
That is, 
\begin{equation}\label{eq:pi_ordering}
\pi_1\le \pi_2\le...\le \pi_J.
\end{equation}
In the following Lemma \ref{lm1}, we present an equivalent version of the IR and IC constraints to simplify Problem \ref{pro3}.
\begin{lemma}
	\label{lm1}
	A contract $\boldsymbol{\phi}=\{(d_{j}, r^L_{j})\}_{j \in \mathcal{J}}$  is feasible (i.e., satisfies IR and IC constraints) if and only if  the contract items satisfy the following three constraints:
	\begin{enumerate}
		\item[a)] $r^L_{J}-\pi_{J}d_{J} \ge 0$;
		\item[b)] $ r^L_{1}\ge ...\ge r^L_{J} \ge 0$ and $ d_{1}\ge ...\ge d_{J}\ge 0$;
		\item[c)] $r^L_{j+1}+\pi_{j}(d_{j}-d_{j+1}) \le r^L_{j} \le r^L_{j+1}+\pi_{j+1}(d_{j}-d_{j+1})$, $j\in \mathcal{J}$.
	\end{enumerate}
\end{lemma}
The proof of Lemma \ref{lm1} is given in Appendix E in the technical report \cite{techtmc}. 
Constraint $(a)$ ensures that each user can get a non-negative payoff by accepting the contract item of type-$J$  users, corresponding to the IR constraints. Both constraints $(b)$ and $(c)$ are related to IC constraints. Constraint $(b)$ shows that the server should request more data from a user type with a lower marginal cost $\pi$ and provide a larger reward in return. Constraint $(c)$ characterizes the relationship between any two neighbor contract items.

Based on Lemma \ref{lm1}, the following Lemma \ref{reward} characterizes the server's optimal learning
rewards for any feasible data size:
\begin{lemma}
	\label{reward}
	For any given data size $\boldsymbol{d}=\{d_j\}_{j\in \mathcal{J}}$ (even if it is not optimal),  the unique optimal     reward for a type $j$ user is:
	\begin{equation}
	\label{31}
	\begin{split}
	&r^{L*}_{j}(\boldsymbol{d})=\\&\hspace{-1mm} \left\{ 
	\begin{array}{rcl}
	&\hspace{-47mm} \pi_{j} d_{j},   & \hspace{-3mm}\rm{if }  $$j = J;$$\\
	&\hspace{-5mm}  \pi_{j} d_{j} + \sum_{m={j+1}}^{J}(\pi_{m}-\pi_{m-1})d_{m},   & \hspace{-3mm}\rm{if }  $$j=1,...,J\hspace{-0.5mm}-\hspace{-0.5mm}1.$$\\
	\end{array} \right.
	\end{split}
	\end{equation}
\end{lemma}
The proof of Lemma \ref{reward} is given in Appendix F in the technical report \cite{techtmc}. Lemma \ref{reward} indicates that all user types except the boundary type $J$ will obtain positive expected payoffs (type-$J$ users receive zero expected payoff), which can be interpreted as the \emph{information rent} in economics due to information asymmetry.

\subsubsection{Optimal Data Sizes in Contract}
\label{subd}
Based on Lemma \ref{reward}, we can
significantly simplify Problem \ref{pro3}  but still   need to   derive the optimal values of $J$  variables $\{d_j\}_{j\in \mathcal{J}}$ under  $J$ constraints $ d_{1}\ge ...\ge d_{J}\ge 0$. 

For the convenience of presentation, we define
\begin{equation}
\hspace{-27.5mm}A_j\triangleq\frac{\varrho I_j(1-p_j+p_jq_j)}{T},
\end{equation}
\begin{equation}
\begin{split}
B_j\triangleq&\gamma I_j\left(p_jq_j\left(\alpha\theta_j+\xi_j\mathbb{E}[\ell_j]\right)+\left(1-p_j\right)\pi_{j}\right)\\&+\sum_{m=1}^{j-1}\gamma I_m(1-p_m)(\pi_j-\pi_{j-1}).
\end{split}
\end{equation}
Based on these two metrics, we first present two special cases of the optimal data sizes, which we call all-independent and all-dependent.
\begin{proposition}
	\label{prop2}
	Two special cases of the optimal data sizes follow: 
	\begin{itemize}
		\item \emph{All-independent.} 		
		If 
		\begin{equation}
		\frac{A_1}{B_1}\ge \frac{A_2}{B_2}\ge ...\ge \frac{A_J}{B_J},
		\end{equation} 
		then the optimal data sizes in the contract are
		\begin{equation}
		\label{ind}
		d_j^*=\sqrt{\frac{A_j}{B_j}},j\in \mathcal{J}.
		\end{equation}
		\item \emph{All-dependent.} 		
		If 
		\begin{equation}
		\label{dependent}
		\frac{\sum_{m\in\mathcal{J}}A_m}{\sum_{m\in\mathcal{J}}B_m}>\frac{\sum_{m=1}^{j}A_m}{\sum_{m=1}^{j}B_m}, \forall j=1,2,...,J-1,
		\end{equation}
	then the optimal data sizes in the contract are
		\begin{equation}
		\label{de}d_j^*=\sqrt{\frac{\sum_{m \in \mathcal{J}}A_m}{\sum_{m \in \mathcal{J}}B_m}},j\in \mathcal{J}.
		\end{equation}
	\end{itemize}
\end{proposition} 
The proof of Proposition \ref{prop2} is given in Appendix G in the technical report \cite{techtmc}. 
The all-independent case means that if  $\{A_j/B_j\}_{j\in\mathcal{J}}$ follow a descending order, then the optimal data size for each type-$j$ user only depends on his own parameters $(A_j,B_j)$. The condition for the all-dependent case means that for any type $j$, there always exists at least one type $m >j$ with  $A_m/B_m$ larger than $A_j/B_j$ (i.e., not in descending order). In this case,  each type's optimal data size depends on all types' parameters $\{(A_j,B_j)\}_{j\in \mathcal{J}}$.


Next, we give an efficient algorithm to compute the optimal data sizes in any possible case based on the insights in Proposition \ref{prop2}.

\begin{theorem}
	\label{thm1}
	For a fixed $J$, there are $2^{J-1}$ possible cases of the optimal data sizes depending on the values of $\{(A_j,B_j)\}_{j\in \mathcal{J}}$. For any given $\{(A_j,B_j)\}_{j\in \mathcal{J}}$, the unique optimal data sizes can be calculated by Algorithm \ref{alg:A}.
\end{theorem}

\setlength{\intextsep}{0pt}
\setlength{\textfloatsep}{0pt}

\begin{algorithm}
	\caption{Optimal data sizes in contract}
	\label{alg:A}
	\SetKwInOut{Input}{Input}\SetKwInOut{Output}{Output}
	\Input  {Parameters $\{(A_j,B_j)\}_{j\in \mathcal{J}}$ indexed based on \eqref{eq:pi_ordering}}
	\Output  {Optimal data sizes $\{(d_j^*)\}_{j\in \mathcal{J}}$}
	
	Initialize $d_j^*\leftarrow\sqrt{\frac{A_j}{B_j}}, j\in \mathcal{J}$;
	
	Find all non-descending  types $\{j:\exists m>j, \frac{A_m}{B_m}>\frac{A_j}{B_j} \text{ or } \exists m<j, \frac{A_m}{B_m}<\frac{A_j}{B_j} \}$;
	
	Put each group of non-descending types that have	adjacent indexes into one auxiliary set $\mathcal{J}_x$;
	
	$X\leftarrow$ the number of these auxiliary sets;\footnote{16}\tcp{i.e., $x\in\{1,2,...,X\}$}

	
	\SetKwFunction{FMain}{check}	
	\For{ $x=1$; $x \le X$; $x++$ }{
		
		\FMain{$\mathcal{J}_x$};\footnote{17}\tcp{divide each auxiliary set $\mathcal{J}_x$ into subsets $\{\mathcal{J}_x^y\}$ that satisfy \eqref{dependent}} 
		

	} 
	
	\SetKwProg{Fn}{Function}{:}{}
	\Fn{\FMain{$\mathcal{J}$}}{
		
		Reindex the types in $\mathcal{J}$ with $1_\mathcal{J},2_\mathcal{J},...,J_\mathcal{J}$.
		
		\If{$|\mathcal{J}|\neq1$ }{
			flag$\leftarrow$1;

			\For{$m=1_\mathcal{J}$ to $(J-1)_\mathcal{J}$}{
				\If{$ \frac{\sum_{j \in \mathcal{J}}A_j}{\sum_{j \in \mathcal{J}}B_j}\le\frac{\sum_{j=1_\mathcal{J}}^{m}A_j}{\sum_{j=1_\mathcal{J}}^{m}B_j}$\tcp{$\mathcal{J}$ does not satisfy \eqref{dependent}} }{
					
					flag$\leftarrow$0;
					
					$d_{j}^*=\sqrt{\frac{\sum_{n=1_\mathcal{J}}^{m}A_n}{\sum_{n=1_\mathcal{J}}^{m}B_n}}, j\in\{1_\mathcal{J},...,m\}$;
					
					$check(\{m+1,...,J_\mathcal{J}\})$; 

					break;}}		
			
			\If{flag=1}{$d_{j}^*=\sqrt{\frac{\sum_{m \in \mathcal{J}}A_m}{\sum_{m \in \mathcal{J}}B_m}}, j\in\mathcal{J}$;\tcp{$\mathcal{J}$  satisfies \eqref{dependent}}}	
		}
	}
\end{algorithm}

\footnotetext[16]{For example, if $\frac{A_1}{B_1}\ge \frac{A_4}{B_4}\ge \frac{A_3}{B_3}\ge \frac{A_2}{B_2} \ge \frac{A_5}{B_5} \ge \frac{A_8}{B_8} \ge \frac{A_6}{B_6}\ge \frac{A_7}{B_7}$, then $X=2$, $\mathcal{J}_1=\{2,3,4\}$, and $\mathcal{J}_2=\{6,7,8\}$.} 

\footnotetext[17]{In the example of footnote 16. If $\frac{A_6}{B_6}\ge \frac{A_7+A_8}{B_7+B_8}$, then $\mathcal{J}_2$ can be divided into two subsets $\{6\}$ and $\{7,8\}$. 
	The optimal data sizes in this example are $d_j^*=\sqrt{\frac{A_j}{B_j}},j=1,5,6$, $d_j^*=\sqrt{\frac{ A_2+A_3+A_4}{ B_2+B_3+B_4}},j=2,3,4$, and  $d_j^*=\sqrt{\frac{ A_7+A_8}{ B_7+B_8}},j=7,8$.}

The proof of Theorem \ref{thm1} is given in Appendix H in the technical report \cite{techtmc}. 
The computation complexity of Algorithm \ref{alg:A} is $\mathcal{O}(\sum_{x=1}^XJ_x)$, which is no larger than $\mathcal{O}(J)$. 
We can interpret Algorithm \ref{alg:A} as    greedily merging non-descending types based on ${A_j}/{B_j}$, so that all merged types have ${\sum_jA_j}/{\sum_jB_j}$ in a descending order.\footnote{In the   example of footnotes 16 and 17,  $\frac{A_1}{B_1}\ge \frac{ A_2+A_3+A_4}{ B_2+B_3+B_4} \ge \frac{A_5}{B_5} \ge  \frac{A_6}{B_6}\ge \frac{ A_7+A_8}{ B_7+B_8}$.} The optimal data sizes of the merged types are the same and follow the dependent form \eqref{de} in Proposition \ref{prop2}, while the optimal data sizes of the not-merged types follow the independent form  \eqref{ind}, as illustrated in footnote 17.

%

\section{Optimal Incentive Mechanism in Unlearning-Forbidden Scenario}
\label{forbidden}
In this section, we first derive the server's optimal incentive mechanism in the unlearning-forbidden scenario in Section \ref{sec1}, then we compare the unlearning-allowed and unlearning-forbidden scenarios based on the server's expected cost and users' expected payoffs in Section \ref{sec2}.
%

\subsection{Server's Optimal Contact Design}
\label{sec1}
Similar to the server's contract design  in the unlearning-allowed scenario (i.e., Problem \ref{pro3}), the server designs the contract to minimize its expected cost in \eqref{uw} under the IR and IC constraints in the unlearning-forbidden scenario:
\begin{problem}
	\label{pro5}
	\begin{equation}
	\begin{split}
	\min \;& \sum_{j' \in \mathcal{J}'}\left(\frac{\varrho I'_{j'}}{Td'_{j'}} +\gamma I'_{j'}r^{L\prime}_{j'}\right),\\
	\rm{s.t.}\; &r^{L\prime}_{j'}-\Pi_{j'}d'_{j'}\ge 0, \forall j'\in \mathcal{J}', \text{ (IR)}	\\
	&r^{L\prime}_{j'}-\Pi_{j'}d'_{j'}\ge r^{L\prime}_{m}-\Pi_{j'}d'_{m}, \forall j', m \in \mathcal{J}',\text{ (IC)}	\\
	\rm{var.}\; &\left\{\left(d'_{j'}, r^{L\prime}_{j'}\right)\right\}_{j' \in \mathcal{J}'},
	\end{split}
	\end{equation}
	where 
	\begin{equation}
	\Pi_{j'}\triangleq\theta_{j'}T+\xi'_{j'}\mathbb{E}[\ell'_{j'}].
	\end{equation}
\end{problem}
For the convenience of presentation, we re-indexed users with $j'$ in an ascending order of $\Pi$, i.e., 
\begin{equation}
\Pi_{1'}\le \Pi_{2'}\le...\le \Pi_{J'},
\end{equation}
and define
\begin{equation}
A'_{j'}\triangleq\frac{\varrho I'_{j'}}{T},
\end{equation}
\begin{equation}
B'_{j'}\triangleq
\gamma\left(\Pi_{j'}\sum_{m=1'}^{j'} I'_m- \Pi_{j-1'} \sum_{m=1'}^{j-1'} I'_m\right).
\end{equation}

After a similar analysis to Section \ref{si}, we obtain the following theorem about the server's optimal contract in the unlearning-forbidden scenario.
\begin{theorem}
	\label{thm2}
	The optimal data sizes $\boldsymbol{d}^{\prime*}$ can be obtained from Theorem \ref{thm1} by substituting $\{(A_j,B_j)\}_{j\in \mathcal{J}}$ with $\{(A'_{j'},B'_{j'})\}_{j'\in \mathcal{J}'}$.
	The optimal rewards 	are
		\begin{equation}
	\label{31wd}
	\begin{split}
	&r^{L\prime*}_{j'}(\boldsymbol{d}^{\prime*})=\\& \left\{ 
	\begin{array}{rcl}
	& \hspace{-50mm}\Pi_{j'} d^{\prime*}_{j'},    &\hspace{-3mm} \rm{if }\; $$j' = J';$$\\
	&  \hspace{-6mm} \Pi_{j'} d^{\prime*}_{j'} + \sum_{m={j+1'}}^{J'}(\Pi_{m}-\Pi_{m-1})d^{\prime*}_{m},   & \hspace{-3mm}\rm{if }\; $$j'=1',...,J-1'.$$\\
	\end{array} \right.
	\end{split}
	\end{equation}
\end{theorem}
The proof of Theorem \ref{thm2} is given in Appendix I in the technical report  \cite{techtmc}.

\subsection{Comparison}
\label{sec2}
In this subsection, we compare the unlearning-allowed and unlearning-forbidden scenarios, to reveal the economic impact of federated unlearning.

Suppose that a type-$j$ user in the unlearning-allowed scenario corresponds to type $j'$ in the unlearning-forbidden scenario. 
For the convenience of presentation, we first introduce the following definitions:
\begin{equation}
\begin{split}
\Delta U_j=&\sum_{m={j+1}}^J(1-p_j)(\pi_m-\pi_{m-1})\sqrt{\frac{\sum_{m\in\mathcal{J}_m}A_m}{\sum_{m\in\mathcal{J}_m}B_m}}\\&-\sum_{m={j+1'}}^{J'}(\Pi_m-\Pi_{m-1})\sqrt{\frac{\sum_{m\in\mathcal{J}_{m}'}A_m'}{\sum_{m\in\mathcal{J}_{m}'}B_m'}},
\end{split}
\end{equation}
\begin{equation}
\begin{split}
\Delta W=&\sum_{j =1}^J\frac{2}{|\mathcal{J}_j|}\sqrt{\left(\sum_{m\in\mathcal{J}_j}A_m\right)\left(\sum_{m\in\mathcal{J}_j}B_m\right)} \\& -\sum_{j' =1'}^{J'}\frac{2}{|\mathcal{J}_{j'}'|}\sqrt{\left(\sum_{m\in\mathcal{J}_{j'}'}A_m'\right)\left(\sum_{m\in\mathcal{J}_{j'}'}B_m'\right)}.
\end{split}
\end{equation}
Given the definitions, we present the comparison results in Proposition \ref{compare}.
 \begin{proposition}
	\label{compare}
	If 
	$\Delta U_j>0$, then the unlearning-allowed scenario makes a type-$j$ user have a larger payoff (i.e., more beneficial) than the unlearning-forbidden scenario. 	
	If 
	$\Delta W<0$, then the unlearning-allowed scenario makes the server have a smaller cost (i.e., more beneficial) than the unlearning-forbidden scenario.
\end{proposition}
The proof of Proposition \ref{compare}  is given in Appendix J in the technical report  \cite{techtmc}.
We can obtain some insights from qualitative analysis:
\begin{itemize}
	\item If users' perceived privacy costs (e.g., $\{\xi'_{j'}\}_{j'\in \mathcal{J}'}$) in the unlearning-forbidden scenario are very large but unlearning costs (e.g., $\lambda$ and $D(\ell)$) are relatively small, then 
	we will have $\Delta U_j<0$ and $\Delta W<0$, i.e., the unlearning-allowed scenario is worse for users but more beneficial to the server.

\item If the unlearning costs are very large but users' perceived privacy costs in the unlearning-forbidden scenario are relatively small, then 
$\Delta U_j>0$ and $\Delta W>0$, i.e.,  the unlearning-allowed scenario is better for users but worse for the server.
\end{itemize}

It is counter-intuitive that users prefer the scenario where they have large costs. We will present detailed illustrations about the preferences of users and the server through simulations in Section \ref{i}.


\begin{figure*}
	\centering
	\begin{minipage}[t]{0.48\linewidth}
		\centering
		\includegraphics[width=2.1 in]{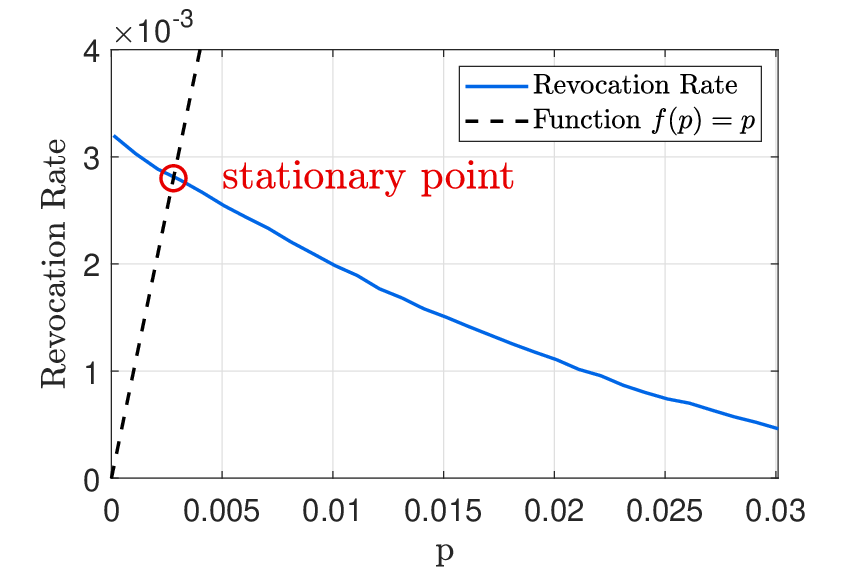}
		\vspace{-4mm}
		\caption{Relationship between the revocation rate $|\mathcal{I}_u^*|/I$ and historical revocation rate $p$.}
		\vspace{-4mm}
		\label{ht22}
	\end{minipage}
	\hspace{0.5mm}
	\begin{minipage}[t]{0.48\linewidth}
		\centering
		\includegraphics[width=2.1 in]{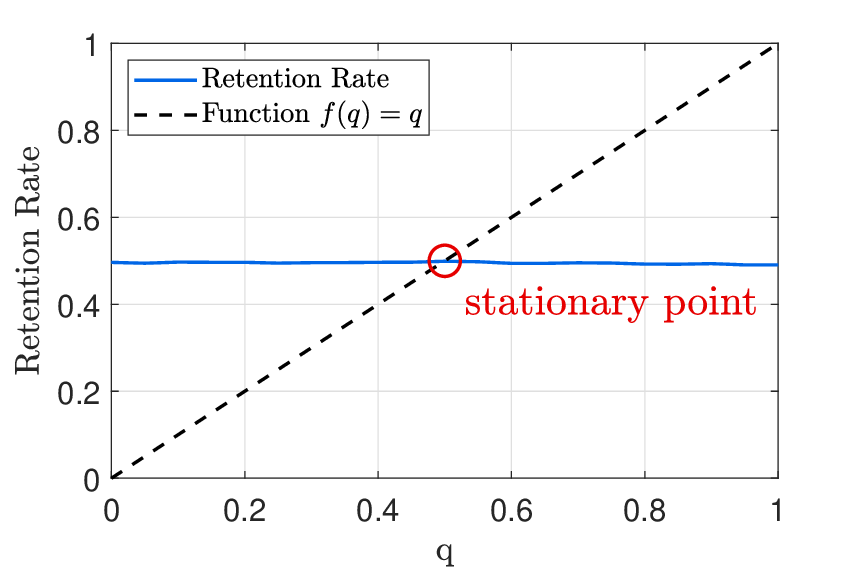}
		\vspace{-4mm}
		\caption{Relationship between the retention rate $|\mathcal{I}_r^*|/|\mathcal{I}_u^*|$ and historical revocation rate $q$.}
		\vspace{-4mm}
		\label{retain1}
	\end{minipage}
\end{figure*}

\section{Simulations}
\label{simulation}
In this section, we use simulations to evaluate the performance of our proposed mechanism. Specifically, in Section \ref{exset}, we specify our experiment setting. In Section \ref{exres}, we validate the optimal strategies of the users and the server in unlearning-allowed and unlearning-forbidden scenarios, and we also compare our mechanism with state-of-the-art benchmarks.

\subsection{Experiment Setting}
\label{exset}
We consider $J=5$ types of users  with marginal training costs  $\boldsymbol{\theta}=[1,4,6,9,10]$, marginal perceived privacy costs  in the unlearning-allowed scenario  $\boldsymbol{\xi}=[0.8,1.7,1.4,$ $2.2,1.2]\times10^3$,\footnote{Different  
	orders of magnitude are to balance different units of users' training costs and privacy costs.}  and marginal perceived privacy costs in the unlearning-forbidden scenario $\boldsymbol{\xi'}=multiplier\cdot\boldsymbol{\xi}$, where $multiplier=8$. Each type has $I_j=I/J=1000$ users. Heterogeneous users'   training losses follow a truncated normal distribution $N(0.5,0.2)$   over the support $[0,1]$, and users' federated Shapley values follow a normal distribution  $N(5\times10^{-5},0.04)$.\footnote{In future work, we will use real-world datasets to calculate users' true training losses and federated Shapley values. The simulation data here can also demonstrate our results. As in Appendix K in the technical report \cite{techtmc}, we further validate that if we change the simulation setting, we will obtain similar experiment results and insights.}  
Users perform $T=100$ rounds of federated learning, and the unlearning rounds coefficient  $\lambda=4$. The server's accuracy loss coefficient  $\varrho=1$ and its  weight on the incentives  $\gamma=10^{-10}$ (to balance different  
units of incentives and model accuracy loss).

We perform experiments to find the appropriate values of historical revocation rate $p$ and retention rate $q$. As shown in Fig.~\ref{ht22} and Fig.~\ref{retain1}, when we  set different values of $p$ and $q$, both the realized revocation rate $|\mathcal{I}_u^*|/I$ and retention rate $|\mathcal{I}_r^*|/|\mathcal{I}_u^*|$ at the equilibrium have a stationary point, i.e., $(2.8\times 10^{-3},2.8\times 10^{-3})$ in Fig.~\ref{ht22} and $(0.5,0.5)$ in Fig.~\ref{retain1}, respectively. 
Therefore, we take the historical revocation rate $p=0.28\%$ and the historical retention rate $q=50\%$ in the following simulations.
\subsection{Experiment Results}
\label{exres}
In Section \ref{i}, we compare the server's expected costs and users' expected payoffs under the optimal contracts in the unlearning-allowed and unlearning-forbidden scenarios. Then, we show users' optimal equilibrium revocation decisions and the server's optimal retention decision in   Section \ref{iii}. Finally, in Section \ref{benchm}, we present the comparison results between our mechanism and two benchmarks. 

\subsubsection{Users' Expected Payoffs and Server's Expected Cost Comparison}
\label{i}
In the following, we show the expected payoff/cost comparison considering three aspects of impact: privacy cost in the unlearning-forbidden scenario, unlearning cost, and training cost.

(i) \emph{Impact of marginal perceived privacy cost in the unlearning-forbidden scenario $\boldsymbol{\xi'}$:}
\begin{figure*}
	\centering
	\begin{minipage}[t]{0.31\linewidth}
		\centering
		\includegraphics[width=2.1 in]{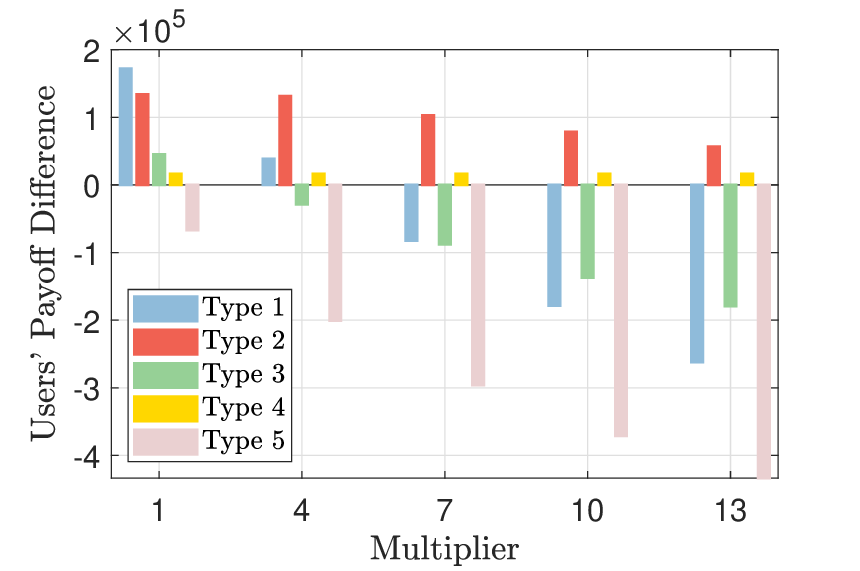}
		\vspace{-4mm}
		\caption{Different types of users' expected payoff difference (unlearning-allowed minus  unlearning-forbidden) versus $\boldsymbol{\xi'}$.}
		\vspace{-3mm}
		\label{ht}
	\end{minipage}
	\hspace{0.5mm}
	\begin{minipage}[t]{0.31\linewidth}
		\centering
		\includegraphics[width=2.1 in]{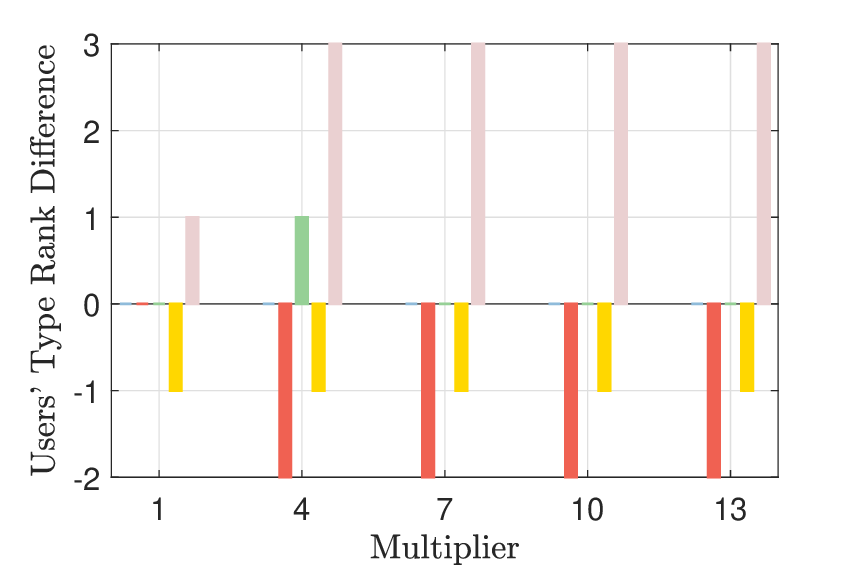}
		\vspace{-4mm}
		\caption{Users' type ranking difference (unlearning-allowed minus  unlearning-forbidden) versus $\boldsymbol{\xi'}$.}
		\vspace{-3mm}
		\label{price}
	\end{minipage}
	\hspace{0.5mm}
	\begin{minipage}[t]{0.34\linewidth}
		\centering
		\includegraphics[width=2.1 in]{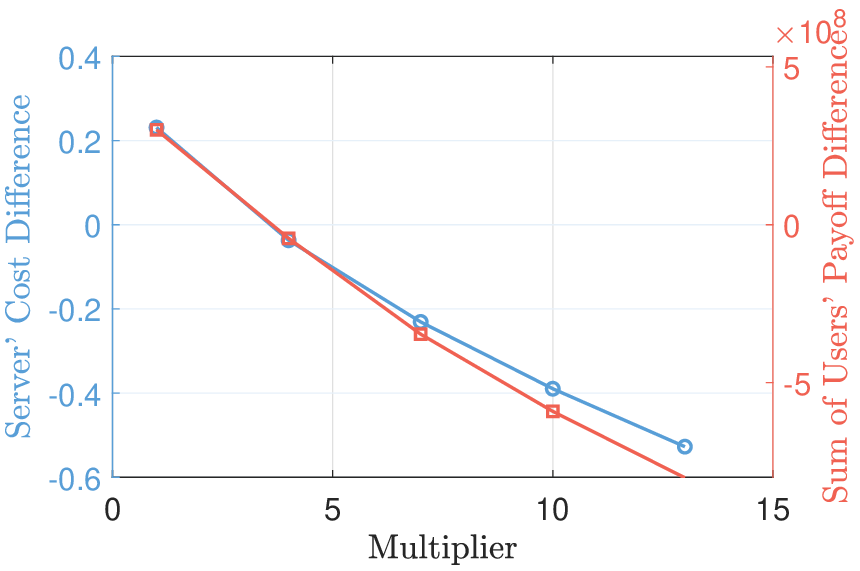}
		\vspace{-4mm}
		\caption{Server's expected cost difference and users' expected total payoff difference (unlearning-allowed minus  unlearning-forbidden) versus $\boldsymbol{\xi'}$.}
		\vspace{-3mm}
		\label{cost}
	\end{minipage}
\end{figure*}


Fig.~\ref{ht} shows different types of users' payoff differences in  unlearning-allowed and unlearning-forbidden scenarios, which indicate their preferences in two scenarios. Different types of users may have different preferences, which are closely related to their type ranking in two scenarios. 

Specifically, as shown in Fig.~\ref{price}, a negative type ranking difference is more likely to lead to a positive payoff difference, i.e., a high ranking  corresponds to a high payoff. This is consistent with the server's optimal rewards in  Lemma \ref{reward} and Theorem \ref{thm2}.

Fig.~\ref{cost} shows users' total expected payoff difference and the server's expected cost difference.  When the perceived privacy cost in the unlearning-forbidden scenario $\boldsymbol{\xi'}$ increases, it is more likely that the unlearning-forbidden scenario is more beneficial to users (i.e., negative payoff difference) but worse for the server (i.e., negative cost difference). The server's preference is straightforward, as a larger $\boldsymbol{\xi'}$ means larger incentive costs for the server. However,   {it is counter-intuitive that users prefer the scenario where they have larger costs, as we may naturally presume that large costs will discourage users' participation.} This is because the server will set the rewards larger than users' costs due to information asymmetry, and the gap (i.e., users' total payoff) increases in users' costs (as indicated in Lemma \ref{reward} and Theorem \ref{thm2}). 

\begin{figure*}
	\centering
	\begin{minipage}[t]{0.48\linewidth}
		\centering
		\includegraphics[width=2.1 in]{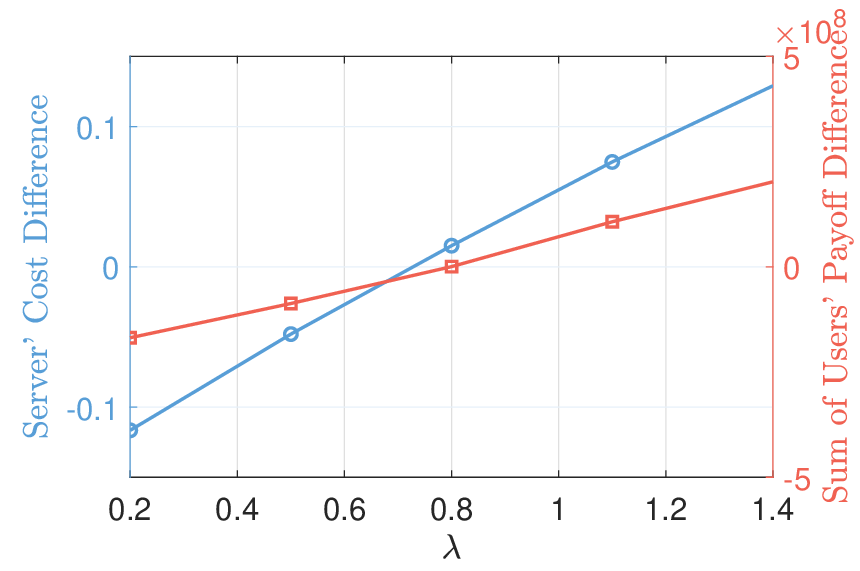}
		\vspace{-3mm}
		\caption{Server's expected cost difference and users' expected total payoff difference (unlearning-allowed minus unlearning-forbidden) versus $\lambda$.}
		\vspace{-4mm}
		\label{cost1}
	\end{minipage}
	\hspace{0.5mm}
	\begin{minipage}[t]{0.48\linewidth}
		\centering
		\includegraphics[width=2.1 in]{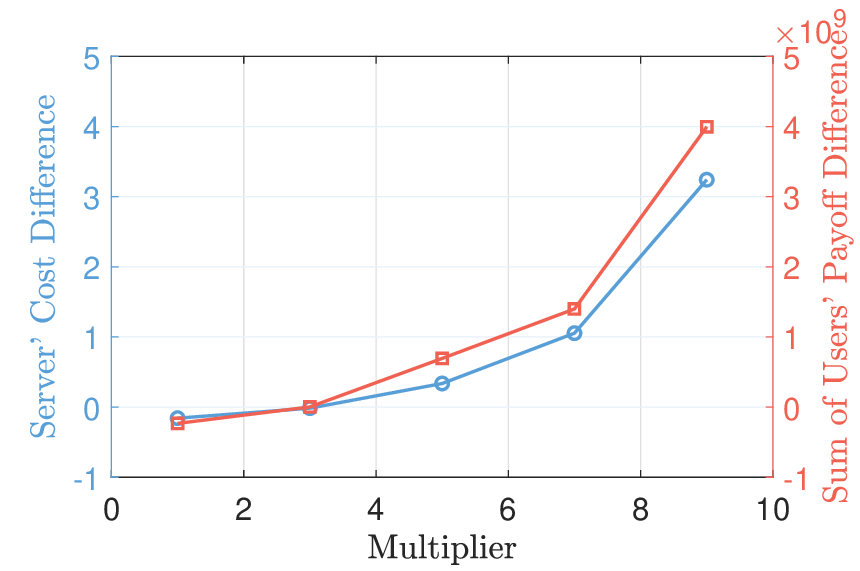}
		\vspace{-3mm}
		\caption{Server's expected cost difference and users' expected total payoff difference (unlearning-allowed minus unlearning-forbidden) versus $\boldsymbol{\theta}$.}
		\vspace{-4mm}
		\label{cost2}
	\end{minipage}
\end{figure*}

(ii) \emph{Impact of unlearning cost:}

Increasing unlearning rounds coefficient $\lambda$ or users' training loss variance $D(\ell)$ will both increase the unlearning cost, so we only simulate the impact of  $\lambda$  here.
As shown in Fig.~\ref{cost1}, when we increase the unlearning cost, it is more likely that the unlearning-allowed scenario is worse for the server but better for users. This is because larger unlearning costs mean larger incentive costs for the server but more rewards for users in the unlearning-allowed scenario. 


Moreover, as shown in Fig.~\ref{cost1}, the server and users' preferences are not always the same (i.e., one positive and the other negative) or different (i.e., the same sign). However, in most cases, they have different preferences.

(iii) \emph{Impact of marginal training cost $\boldsymbol{\theta}$:}

 We increase the value of the marginal training cost $\boldsymbol{\theta}$ by multiplying by a multiplier. 
As shown in Fig.~\ref{cost2}, the insights are similar to that of unlearning cost. The training cost $\boldsymbol{\theta}$ affects both learning cost and unlearning cost. As both scenarios have learning costs, increasing $\boldsymbol{\theta}$ is similar to the effect of increasing the unlearning cost. 

\begin{figure*}
	\centering
	\begin{minipage}[t]{0.48\linewidth}
		\centering
		\includegraphics[width=2.3 in]{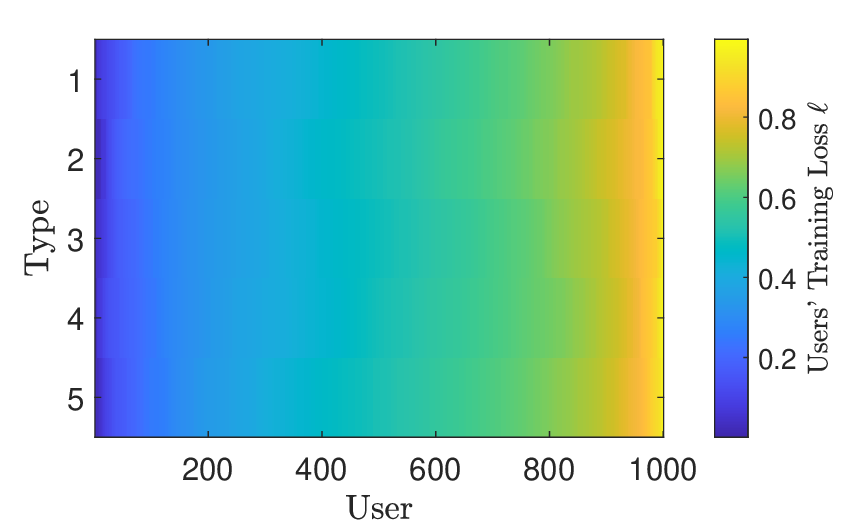}
		\vspace{-4mm}
		\caption{Users' training losses $\{\ell_i\}_{i\in \mathcal{I}}$.}
		\vspace{-3mm}
		\label{price22}
	\end{minipage}
	\hspace{0.5mm}
	\begin{minipage}[t]{0.48\linewidth}
		\centering
		\includegraphics[width=2.3 in]{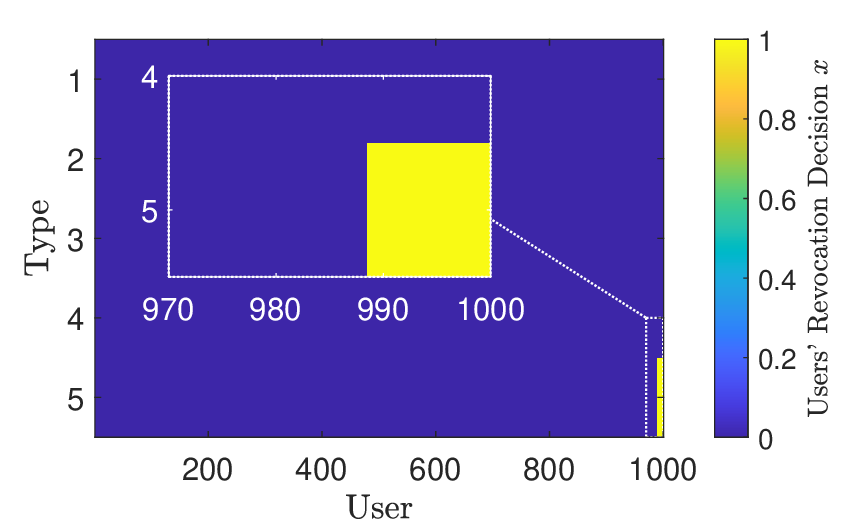}
		\vspace{-4mm}
		\caption{Users' optimal revocation decisions $\{x_i^*\}_{i\in \mathcal{I}}$. 
		}
		\vspace{-3mm}
		\label{cost22}
	\end{minipage}
\end{figure*}

\begin{figure*}
	\centering
	\begin{minipage}[t]{0.33\linewidth}
		\centering
		\includegraphics[width=2.26 in]{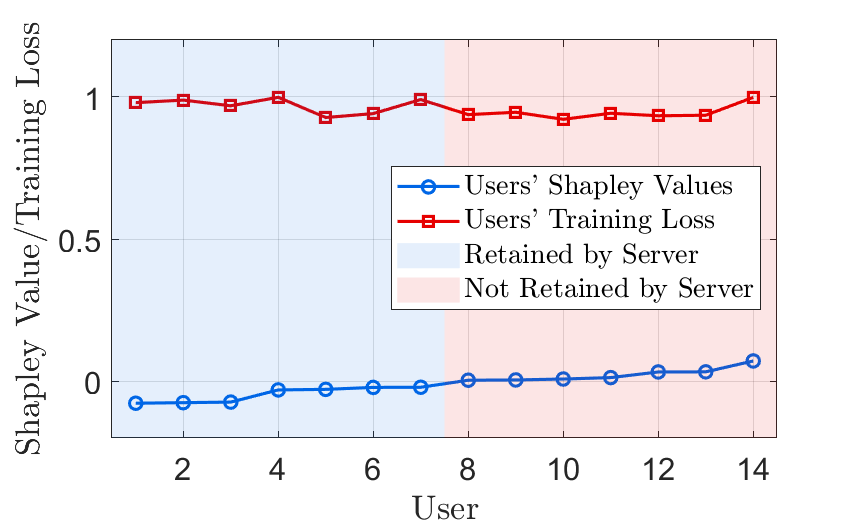}
		\vspace{-4mm}
		\caption{Server's optimal retention decisions $\mathcal{I}_r^*$. 
		}
		\vspace{-5mm}
		\label{retain2}
	\end{minipage}
	\hspace{0.5mm}
	\begin{minipage}[t]{0.32\linewidth}
		\centering
		\includegraphics[width=2.1 in]{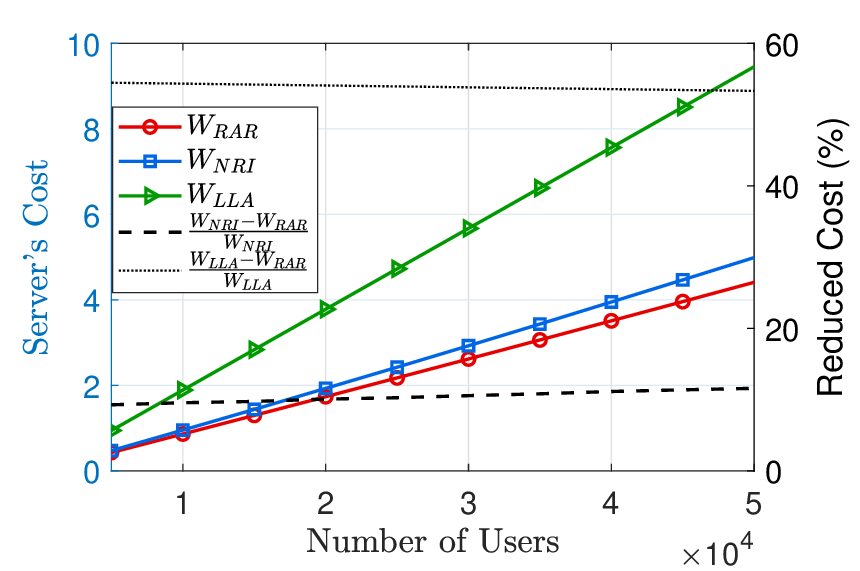}
		\vspace{-4mm}
		\caption{Server's cost comparison of NRI, LLA, and RAR.}
		\vspace{-5mm}
		\label{bench}
	\end{minipage}
	\hspace{0.5mm}
	\begin{minipage}[t]{0.32\linewidth}
		\centering
		\includegraphics[width=2.1 in]{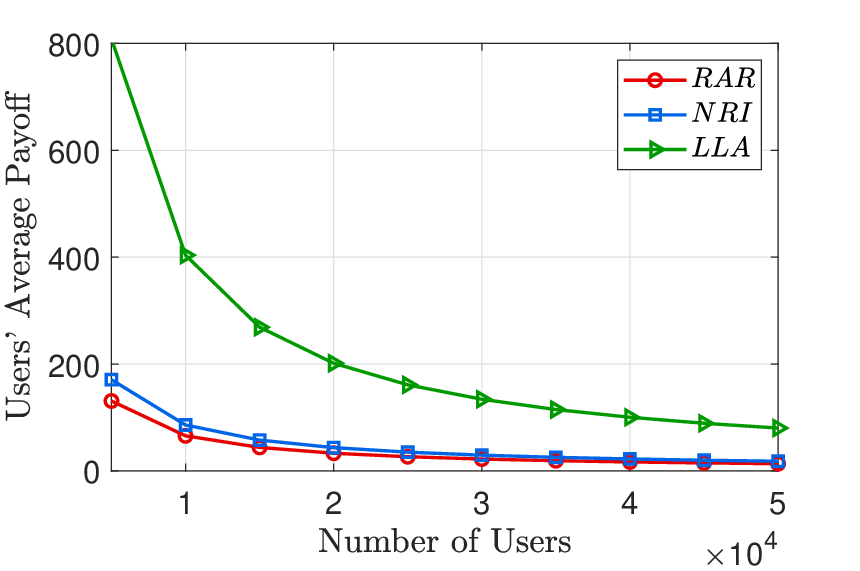}
		\vspace{-4mm}
		\caption{Users' average payoff comparison of NRI, LLA, and RAR.}
		\vspace{-5mm}
		\label{payave}
	\end{minipage}
\end{figure*}

\subsubsection{Users' Revocation Decisions and Server's Retention Decision}
\label{iii}
As shown in Fig.~\ref{price22}, we rank each type of users in ascending order of their training losses for the convenience of presenting insights.

Fig.~\ref{cost22} shows that at the equilibrium, users with larger aggregated marginal costs $\pi$ (i.e., type 5) and training losses $\ell$ (i.e., user 986-1000) are more likely to revoke their data. This is because (i) users with larger costs receive smaller learning incentives from the server in the contract (Lemma \ref{reward}); (ii) they do not know their high training losses before federated learning and their realized privacy costs (training losses) significantly exceed their expectations.

Fig.~\ref{retain2} illustrates the server's optimal retention decision. We rank the users who want to revoke their data in ascending order of their federated Shapley values $\{v_i\}_{i\in \mathcal{I}_u}$. Users with smaller federated Shapley values are more likely to be retained by the server,  as smaller Shapley values represent larger contributions to the global model accuracy.  
Users with smaller training losses have lower privacy costs and may require fewer incentives from the server, compared to users with larger losses. However, Fig.~\ref{retain2} shows that the server does not necessarily retain users with smaller training losses. This is because given a fixed set of users, reducing the total training losses of retained users means increasing the total losses of leaving users, resulting in higher unlearning costs (Proposition \ref{lm}).



\subsubsection{Comparison with Benchmarks}
\label{benchm}
We compare our incentive mechanism with two benchmarks to evaluate the performance. 
\begin{itemize}
	\item No Retention Incentive (NRI): the server does not retain users who want to revoke their data.
	\item Limited Look Ahead (LLA) (adapted from \cite{ding2020optimal}): the server 
	first optimizes the incentive mechanism for federated learning without considering the unlearning part, and then designs the retention incentive in unlearning (i.e., separate optimization).
	\item Our proposed incentive mechanism (RAR): the server is \underline{R}ational in jointly optimizing both federated learning and unlearning  \underline{A}nd designs \underline{R}etention incentive to retain valuable leaving users.
\end{itemize}

Fig.~\ref{bench} shows the server's costs in the three mechanisms under different numbers of users. Our proposed RAR reduces the server's cost by around 53.91\% (black dotted line) compared with LLA. 
The reduced cost of RAR compared with NRI can reach  11.59\%   (black dashed line) and will increase in the number of users, as the server retains more valuable users when the number of users increases. 
Therefore, it is beneficial for the server to retain valuable leaving users and make joint optimization of federated learning and unlearning incentive mechanisms.
As the objective of our incentive mechanism design is to minimize the server's cost, the server's cost reduction is at the expense of users' payoffs (as shown in Fig.~\ref{payave}).

\section{Conclusion}
\label{conclusion}
To the best of our knowledge, this paper is the first study to focus on the important issue of incentive design for federated learning and unlearning. 
We derive theoretical bounds on the global model optimality gap and the number of communication rounds of natural federated unlearning, based on Scaffold and FedAvg algorithms. 
Our approach tackles a challenging problem in incentive design, by summarizing users' multi-dimensional heterogeneity into one-dimensional metrics and developing an efficient algorithm for an exponentially large number of possible cases.
We compare the unlearning-forbidden and unlearning-allowed scenarios in terms of users' payoffs and the server's cost. Counter-intuitively, users usually prefer the scenario where they have larger costs. This is because the server will give them even higher incentives than their costs due to information asymmetry.
We also identify what types of users will leave the system or be retained by the server. 
The experiments demonstrate the superior performance of our proposed incentive mechanism and the benefits of unlearning incentives for retaining leaving users.
We will design incentive mechanisms for federated unlearning to maximize social welfare in future work.

\bibliographystyle{IEEEtran}
\bibliography{ref} 


\end{document}